\documentclass[elsart,10pt,onecolumn]{elsarticle}

\usepackage{graphics,graphicx}
\usepackage{amsmath,amsfonts,amssymb}
\usepackage{algorithm}
\usepackage{algpseudocode}

\newcommand{\Xb}{\mathbf{X}}
\newcommand{\Xm}{\mathbb{X}}

\newcommand{\Zb}{\mathbf{Z}}

\newcommand{\Fb}{\mathbf{F}}
\newcommand{\Ub}{\mathbf{U}}
\newcommand{\Qb}{\mathbf{Q}}
\newcommand{\Ib}{\mathbf{I}}

\newcommand{\xb}{\mathbf{x}}
\newcommand{\vb}{\mathbf{v}}

\newcommand{\wb}{\mathbf{w}}

\newcommand{\zb}{\mathbf{z}}
\newcommand{\fb}{\mathbf{f}}
\newcommand{\hb}{\mathbf{h}}

\newcommand{\sD}{\text{\tiny D}}
\newcommand{\sS}{\text{\tiny S}}

\newcommand{\Gammab}{\mbox{\boldmath$\Gamma$}}
\newcommand{\thB}{\mbox{\boldmath$\theta$}}

\newcommand{\ThB}{\mbox{\boldmath$\Theta$}}

\journal{Information Fusion}

\begin{document}

\begin{frontmatter}

\title{An Overview of Particle Methods for Random Finite Set Models}

\author[RMIT]{Branko Ristic\corref{cor1}}
\cortext[cor1]{Corresponding author: B. Ristic, RMIT University, School of Electrical and Computer Engineering, Rm 10.8.14, 376-392 Swanston street,  Melbourne,  VIC 3000, Australia; email: branko.ristic@rmit.edu.au; tel: +61 3 9925 3768}
\address[RMIT]{RMIT University, Australia}
\address[CU]{Curtin University, Australia}
\address[CF]{
Universit\`{a} di Firenze, DINFO, Florence, Italy}

\author[CU]{Michael Beard}
\author[CF]{Claudio Fantacci}

\begin{abstract}
This overview paper describes the particle methods developed for the
implementation of the a class of Bayes filters formulated using the random finite set formalism. It is primarily intended for the readership already familiar with the
particle methods in the context of the standard Bayes filter. The focus in on the Bernoulli particle filter, the probability hypothesis density (PHD) particle filter and the generalised labelled multi-Bernoulli (GLMB) particle filter. The performance of the described filters is demonstrated in the context of bearings-only target tracking application.
\end{abstract}


\end{frontmatter}
\newpage
\tableofcontents
\section*{Notation}
{
\begin{tabular}{ l l }
  $\mathcal{X}$ & the single object state space\\
  $\mathcal{F}(\mathcal{X})$ & the multiple object state space\\
  $\mathcal{L}$ & the space of labels\\
  $\xb$  & the state of a single object (a random vector) \\
  $\Xb$ & the state of multiple-objects (a random finite set, RFS) \\
  $\mathbb{X}$ & the state of multiple labelled objects (a labelled
  RFS)\\
  $\mathcal{Z}$ & measurement space \\
  $\zb$ & a measurement of a single object (a random vector)\\
  $\Zb$ & a detector output measurement (a RFS)\\
  $k$ & discrete-time index\\
  $p(\xb)$ & a probability density function (PDF) of  $\xb\in\mathcal{X}$\\
  $\tilde{p}(\xb,\ell)$ & a PDF of a labelled random vector $(\xb,\ell)\in \mathcal{X} \times \mathcal{L}$\\
  $f(\Xb)$ & a PDF of an RFS variable  $\Xb\in\mathcal{F}(\mathcal{X})$\\
  $\tilde{f}(\mathbb{X})$ & a PDF of a labelled RFS variable  $\mathbb{X}\in(\mathcal{F}(\mathcal{X})\times\mathcal{L})$\\
\end{tabular}
}

\section{Introduction}

In many areas of science and engineering there is a need to infer
the behaviour of a stochastic dynamic system, using its partial and
indirect observations. By combining (typically nonlinear)
mathematical models of system evolution and sensor measurements, one
can formulate the optimal sequential estimator in the Bayesian
framework. This estimator, commonly referred to as the Bayes-optimal
(or simply Bayes) filter, provides a recursive formula for the
complete probabilistic characterisation of the dynamic system in the
form of a time-varying posterior probability density of its state \cite{jazwinski_70}.

For most nonlinear/non-Gaussian formulations, analytic closed-from
solutions of the Bayes filter are intractable. Practical solutions,
therefore, need to be based on approximations. Particle filters are
a class of Monte Carlo simulation based methods which can provide
very accurate approximations of the Bayes filter. Despite being
computationally expensive, particle filters have become universally
popular, primarily due to their accuracy, relatively simple
implementation and the ever increasing speed of computers. As a result
of their widespread application, a few good tutorials and books have
been published on the subject of particle filters \cite{smcbook},
\cite{pf_tute}, \cite{djuric_pf_03}, \cite{pfbook}, \cite{cape_07},
\cite{doucet_15_years}.

Particle filters have been introduced and traditionally applied as
the approximate solutions of the {\em standard} Bayes filter,
formulated during the 1960s \cite{jazwinski_70}  under the following
assumptions: (i) the stochastic dynamic system (object, phenomenon)
is permanently active (or present); (ii) observations are noisy, but
collected with  perfect detection (i.e. there are no false or missed
detections). All the aforementioned tutorials and books discuss the
particle filters in this context only. However, in many practical
applications, one may have to deal with multiple stochastic dynamic
systems (objects), which can be simultaneously active (present), and
which can randomly switch {\em on} and  {\em off}
(appear/disappear). In addition, perfect detection using
surveillance sensors (e.g. radar, sonar, video cameras) is rarely
possible \cite{kay1998}. Until recently, particle filters have been applied to this
class of problems using a clever combination of Bayesian estimation
theory with ad-hoc logic. However, the recent advances in Bayesian
estimation using {\em random finite set} (RFS) models
\cite{mahler_14} resulted in elegant and rigorous mathematical
formulations of the Bayes-optimal and principled Bayes-suboptimal
filters, applicable to multiple interacting on/off switching systems
with possibly imperfect detection of measurements.

This overview paper describes the particle methods developed for the
implementation of the new class of RFS-Bayes filters. It is
primarily intended for the readership already familiar with the
particle methods in the context of the standard Bayes filter.  One
of the most popular and convincing applications of particle
filters, versus standard approximation methods, such as the Extended
Kalman filter (EKF) \cite{barshalom_et_al_01}
and unscented Kalman filter (UKF) \cite{julier2004unscented}, has been for
bearings-only tracking problems \cite[Ch.6]{pfbook}. Hence, this
application has been chosen to demonstrate throughout the paper
different RFS-Bayes particle filters and their performance. The
paper is organised as follows. Sec. \ref{s:2A} reviews the particle method for
the standard Bayes filter. The elements of mathematics for random finite set models are
presented in Sec. \ref{s:RFS}. The particle method for the RFS Bayes-optimal filter and its special case, the Bernoulli filter, are discussed in Sec. \ref{s:4}. A multi-target particle filter, referred to as the PHD particle filter, is presented in Sec. \ref{s:phd}. The labelled RFS Bayes tracking filters implemented using the particle method are discussed in Sec. \ref{s:lrfs}. The summary and  pointers to advanced research topics are given in Sec. \ref{s:sum}.


\section{Particle methods for the standard Bayes filter}
\label{s:2A}

In order to familiarise with the notation, let us start with a quick
review of the standard Bayes filter and the corresponding particle
methods. The problem is by no means simple and it is still an
active and highly relevant research topic.

\subsection{Problem formulation and the standard Bayes filter}
\label{s:2}

Suppose the state vector $\xb_k\in\mathcal{X}$ provides the complete
specification of the state of a dynamic system (object, phenomenon)
at time $t_k$. Here $\mathcal{X}\subseteq\mathcal{R}^{n_x}$ is the
state space, while $k$ is the discrete-time index corresponding to
$t_k$. Let us adopt the discrete-time additive-noise formulation,
specified by two equations:
\begin{eqnarray}
        \xb_{k}= &  \fb_{k-1}(\xb_{k-1})+ \vb_{k-1},   \label{e:dyn_model}\\
        \zb_{k}= &  \hb_k(\xb_{k}) + \wb_{k},    \label{e:meas_model}
\end{eqnarray}
referred to as the {\em dynamics equation} and the {\em measurement
equation}, respectively. Function $\fb_{k-1}:\mathcal{R}^{n_x}
\rightarrow \mathcal{R}^{n_x}$ in (\ref{e:dyn_model}) is a nonlinear
transition function defining the temporal evolution of the state
vector as a first-order Markov process. Random disturbances
$\vb_{k}\in\mathcal{R}^{n_x}$, also known as process noise,    are
assumed to be independent identically distributed~(IID) according to
the probability density function (PDF) $p_{\vb}$. Function
$\hb_k:\mathcal{R}^{n_x} \rightarrow \mathcal{R}^{n_z}$ in (\ref{e:meas_model}) defines the
relationship between the state $\xb_k$ and the measurement
$\zb_{k}\in\mathcal{Z}$, where
$\mathcal{Z}\subseteq\mathcal{R}^{n_z}$ is the measurement space.
Random disturbances $\wb_k\in\mathcal{R}^{n_z}$, also known as
measurement noise, are assumed independent of $\vb_k$, and modelled
as an IID process with the PDF $p_{\wb}$. Typically $n_{z} < n_x$,
giving rise to the term {\em partial observations} of the system.

In the formulation specified by
(\ref{e:dyn_model})-(\ref{e:meas_model}), the functions $\fb_k$ and
$\hb_k$, the probability distributions $p_{\vb}$ and $p_{\wb}$, and
the PDF of the state vector at initial time $k=0$, (i.e.
$p_0(\xb_0)$), are all assumed known. Equations (\ref{e:dyn_model})
and (\ref{e:meas_model}) effectively define two probability
functions: the {\em transitional density} $
\pi_{k|k-1}(\xb_k|\xb_{k-1})= p_{\vb}\left(\xb_k-
\fb_{k-1}(\xb_{k-1}) \right)$ and the {\em likelihood function}
$g_k(\zb_k|\xb_k) = p_{\wb}\left(\zb_k-\hb_k(\xb_k)\right)$. The
problem is to compute recursively the posterior PDF of the state,
denoted as $p_{k|k}(\xb_k|\zb_{1:k})$ at discrete-time $k$, where
the notation $\zb_{1:k}$ stands for the sequence
$\zb_{1},\zb_2,\cdots,\zb_k$.

The solution is usually presented as a two step procedure. Let
$p_{k-1|k-1}(\xb_{k-1}|\zb_{1:k-1})$ denote the posterior PDF at
$k-1$. The first step {\em predicts} the density of the state to
time $k$ 
via the
Chapman-Kolmogorov equation \cite{jazwinski_70}:
\begin{equation}
p_{k|k-1}(\xb_k|\zb_{1:k-1}) \hspace{-1.mm}= \hspace{-1.5mm}\int
\pi_{k|k-1}(\xb_k|\xb')p(\xb'|\zb_{1:k-1}) d\xb'. \label{e:pred_eq}
\end{equation}
The second step applies Bayes rule to {\em update}
$p(\xb_k|\zb_{1:k-1})$ using  measurement $\zb_k$:
\begin{equation}
p_{k|k}(\xb_k|\zb_{1:k}) = \frac{g_k(\zb_k|\xb_k)\,
p_{k|k-1}(\xb_k|\zb_{1:k-1})}{\int g_k(\zb_k|\xb)
p_{k|k-1}(\xb|\zb_{1:k-1}) d\xb}. \label{e:upd_eq}
\end{equation}
Knowing the posterior $p_{k|k}(\xb_k|\zb_{1:k})$, one can compute a
point estimate of the state $\hat{\xb}_k$ (e.g. as the mean
or the mode of the posterior) and a confidence (or credible)
interval.

The closed-form analytic solution to
(\ref{e:pred_eq})-(\ref{e:upd_eq}) can be found only in some special
cases. One important case is when  $\fb_k$ and $\hb_k$ are linear
functions and PDFs $p_{\vb}$, $p_{\wb}$ and $p_0$ are Gaussian; the
solution in this case is the Kalman filter. In general, however,
stochastic filtering via (\ref{e:pred_eq})-(\ref{e:upd_eq}) can be
solved only numerically. Many algorithms have been proposed for
this purpose, including analytic approximations (e.g. Extended
Kalman filter and its variants), grid-based methods (where the
posterior PDF is evaluated at a finite and fixed set of points),
Gaussian sum filters (where the posterior PDF is approximated by a
Gaussian mixture), unscented transforms \cite{julier2004unscented} and particle filters
\cite{pfbook}, \cite{challa}.

\subsection{A primer on the particle method}
\label{s:primer}

Suppose the posterior density at discrete-time $k-1$  is
approximated by a set of random samples (particles)
$\{w^{(i)}_{k-1},\xb^{(i)}_{k-1}\}_{1\leq i \leq N}$, where
$\xb^{(i)}_{k-1}$ is the state of particle $i$ and $w^{(i)}_{k-1}$
is its weight. The weights are normalized, that is $\sum_{i=1}^N
w^{(i)}_{k-1} =1$. This approximation of the posterior improves as
$N\rightarrow \infty$.
Given $\{w^{(i)}_{k-1},\xb^{(i)}_{k-1}\}_{1\leq i \leq N}$ and using
the measurement $\zb_k$ at time $k$, the key question is how to form the
particle approximation of the posterior at $k$, i.e.
$p_{k|k}(\xb_k|\zb_{1:k})$, denoted  $\{w^{(i)}_k,
\xb^{(i)}_k\}_{1\leq i \leq N}$.

The computation of the weights and particles at time $k$ is based on
the concept of importance sampling \cite{robert_casella99}. Let us
introduce a proposal or importance density
$q_k(\xb_k|\xb_{k-1},\zb_k)$, whose support contains the support of
the posterior PDF at time $k$. Then the (preliminary) particles at
time $k$ are drawn from the importance density:
\begin{equation}
\tilde{\xb}^{(i)}_k  \sim  q_k(\xb_k|\xb^{(i)}_{k-1},\zb_k),
\label{e:pf1}
\end{equation}
whose weights are computed as follows:
\begin{eqnarray}
 \tilde{w}^{(i)}_k & = & w^{(i)}_{k-1}
\frac{g_k(\zb_k|\tilde{\xb}^{(i)}_k)\,\pi_{k|k-1}(\tilde{\xb}^{(i)}_k|\xb^{(i)}_{k-1})}{q_k(\tilde{\xb}^{(i)}_{k}|\xb^{(i)}_{k-1},\zb_k)}
\label{e:unnw}\\
 w^{(i)}_k & = &
\frac{\tilde{w}^{(i)}_k}{\sum_{j=1}^N \tilde{w}^{(j)}_k}
\label{e:pf3}
\end{eqnarray}
for $i=1,\dots,N$. This recursive procedure starts at time $k=0$ by
sampling $N$ times from the initial PDF $p_0$.

The described particle method, also known as sequential importance
sampling (SIS), inevitably fails after many iterations, because all
particle weights, except a few, become zero (a poor approximation of
the posterior PDF due to particle degeneracy). The collapse of the
SIS scheme can be prevented by resampling the particles. The
resampling step chooses $N$ particles from $\{w^{(i)}_k,
\tilde{\xb}^{(i)}_k\}_{1\leq i \leq N}$, where the selection of
particles is based on their weights: the probability of particle $i$
being selected during resampling equals $w^{(i)}_k$. After
resampling, all particle weights are equal to $1/N$. While
resampling avoids degeneracy of particles, it leads to the loss of
diversity among the particles, because the particles with large
weights are selected (repeated) many times. In order to increase the
particle diversity, it is usually recommended to perform a Markov
chain Monte Carlo (MCMC) move \index{MCMC move} step after
resampling \cite{gilks}, \cite{pfbook}, \cite{cape_07}.

The choice of the importance density $q_k$ plays an important role
in the implementation of the particle filter (PF). The simplest
choice is to select $q_k$ as the transitional density, i.e.
$q_k\equiv\pi_{k|k-1}$. This PF, referred to as the bootstrap filter
\cite{gordon_etal_93}, can lead to poor performance because many of
the particles could be sampled from the region of the state space
which is not in the support of the posterior (and therefore
wasted). A better strategy is to use the information contained in
the latest measurement $\zb_k$ in the design  of the importance
density. Research into good importance densities have resulted in
many versions of the particle filter, such as: the optimal
importance density (OID) PF \cite{doucetetal99}, the auxiliary PF
\cite{PITTSHEPHARD}, the local-linearisation PF \cite{doucetetal99},
\cite{merwe}, exact particle flow nonlinear filters \cite{daum_10},
particle filters with progressive correction or tempering
\cite{musso_et_al_00} and particle filters using Laplace
approximation \cite{paul_quang_15}.

The bootstrap filter, being the simplest and possibly the most
popular PF, is described in more detail and its  pseudo code (for a
single processing cycle at time $k$) is given in Alg. \ref{a:1}.
Note that, since resampling is performed at the end of every cycle,
there is no need to input/output particle weights. Recall also that
after resampling it is recommended to carry out an MCMC move step.

\begin{algorithm}[ht]
\caption{ Pseudo-code of the bootstrap filter } { \footnotesize
\begin{algorithmic}[1]
\Function{Bootstrap Filter}{} \State Input:
$\{\xb^{(i)}_{k-1}\}_{1\leq i \leq N}$; $\zb_k$
 \For{$i=1,\dots,N$}
   \State Draw a sample: $\tilde{\xb}^{(i)}_k \sim
\pi_{k|k-1}(\xb_k|\xb^{(i)}_{k-1})$ \State Calculate weight
$\tilde{w}^{(i)}_k =  g_k(\zb_k|\tilde{\xb}_k^{(i)})$ \EndFor
\State $w_k^{(i)} = \tilde{w}_k^{(i)}/\sum_{j=1}^N
\tilde{w}_k^{(j)}$, for  $i=1,\dots,N$
\For {$i=1,\dots,N$} \Comment (Resampling) \State Select index
$j^i\in\{1,\dots,N\}$ with probability $w_k^{(i)}$ \State
 $\xb^{(i)}_k = \tilde{\xb}^{(j_i)}_{k}$ \EndFor
\State Apply MCMC move and output $\{\xb^{(i)}_{k}\}_{1 \leq i \leq
N}$ \EndFunction
\end{algorithmic}
%
%
} \label{a:1}
\end{algorithm}

A point estimate of the state at time $k$, denoted $\hat{\xb}_k$,
can be computed from the particle approximation
$\{w^{(i)}_k,\xb^{(i)}_{k}\}_{1\leq i \leq N}$ of the true posterior
$p_{k|k}(\xb_k|\zb_{1:k})$ either: (a) as the weighted mean of the
particles, referred to as the expected a posterior (EAP) estimate,
or (b) as the mode of the density estimated from the particles
(using for example the kernel density estimation method, see
\cite{silverman_86}), referred to as the maximum a posterior (MAP)
estimate.

\subsection{Calibration of system parameters}
\label{s:calib}

In many practical applications of nonlinear stochastic filtering,
the transitional density and/or the likelihood function are
dependent on a static parameter vector $\thB\in
\mathbf{\Theta}\subseteq \mathcal{R}^{n_\theta}$. This is indicated
in notation as $\pi_{k|k-1}(\xb_k|\xb_{k-1},\thB)$ and
$g_k(\zb_k|\xb_k,\thB)$. The problem is to estimate the posterior
density $p(\thB|\zb_{1:k})$ (i.e. to calibrate the system) given its
prior $p(\thB)$ and observations $z_{1:k}$.

Note that according to Bayes rule $p(\thB|\zb_{1:k})\propto \varrho(\zb_{1:k}|\thB)\,p(\thB)$, where
$\varrho(\zb_{1:k}|\thB)$ is the likelihood function.
If we knew $\varrho(\zb_{1:k}|\thB)$, we could apply standard Bayesian parameter estimation
techniques (e.g. MCMC, population Monte Carlo
\cite{robert_casella99}). In general, however, $\varrho(\zb_{1:k}|\thB)$ cannot be expressed in closed-form. One natural option is to augment the state vector and
carry out sequential Bayesian estimation on the joint space
$\mathcal{X} \times \ThB$. Unfortunately, this is not a good idea,
because the direct use of particle filtering in this case is
inefficient \cite{cape_07} due to the absence of stochastic
evolution for $\thB$ (i.e. the particles in $\ThB$ space are sampled
effectively only once).

The key idea of particle MCMC methods is to estimate the likelihood
$\varrho(\zb_{1:k}|\thB)$ using the particle filter \cite{pMCMC}. Note
first the following decomposition of the likelihood:
\begin{equation}
\varrho(\zb_{1:k}|\thB) =
\varrho(\zb_1|\thB)\prod_{t=2}^{k}\varrho(\zb_{t}|\zb_{1:t-1},\thB)
\label{e:factor}
\end{equation}
where the terms $\varrho(\zb_{t}|\zb_{1:t-1},\thB)$ can be estimated
from the un-normalised weights of the particles, see
(\ref{e:unnw}),  at time step $t$ \cite{cape_07}:
\begin{equation}
\varrho(\zb_{t}|\zb_{1:t-1},\thB) = \int g_t(\zb_t|\xb)
p_{t|t-1}(\xb|\zb_{1:t-1}) d\xb \approx \sum_{1\leq i \leq N}
\tilde{w}_t^{(i,\thB)}.
\end{equation}
Using for example the Metropolis-Hasting MCMC algorithm,  the parameter
vector values $\thB^*$ are iteratively generated from the proposal distribution $q(\thB|\thB')$ in order to be accepted or rejected. Note that for each proposed $\thB^*$, one has
to run the PF  to estimate its likelihood $\varrho(\zb_{1:k}|\thB^*)$.

Suppose next that the parameter vector is also time varying. For
example, its evolution may also be modeled by a first-order Markov
process. In this case, although the inference needs to be made on
the joint space $\mathcal{X} \times \ThB$, the particle filter can
be executed on sub-spaces $\ThB$ and $\mathcal{X}|\ThB$ because the
posterior PDF can be factorized as $p_{k|k}(\xb_k,\thB_k|\zb_{1:k})
= p_{k|k}(\xb_k|\thB_k,\zb_{1:k})p_{k|k}(\thB_k|\zb_{1:k})$. In some
applications, the conditional posterior
$p_{k|k}(\xb_k|\thB_k,\zb_{1:k})$ is analytically tractable (e.g.
linear/Gaussian case), which leads to the Rao-Blackwellised
formulation of the marginalised PF \cite{cape_07}, \cite{schon_05}.

\subsection{Demonstration: Bearings-only filtering}
\label{s:bot1}
 The problem of bearings-only filtering (or tracking)
arises in a variety of important applications, including submarine
tracking, using a passive sonar, and aircraft surveillance, using a
radar in a passive mode \cite{pfbook}. The objective is to sequentially estimate
the kinematics of a moving object using noise-corrupted bearing
measurements. Two features of the problem are noteworthy
\cite{pfbook},\cite{SRF_07}: (a) the observation platform needs to
manoeuvre in order to estimate the target range; (b) the problem
becomes particulary difficult during the time steps when the
bearings-rate is high. The particle filter, as a universal method
for nonlinear filtering, has been demonstrated to outperform both
EKF and UKF in this application\footnote{Notice that the shifted
Rayleigh filter \cite{SRF_07}, which is also a very accurate
bearings-only filter, is not a universal nonlinear filter.}
\cite{pfbook},\cite{SRF_07}.

Let us adopt the state vector of the moving object as
\begin{equation}
\xb^m_k = \left[\begin{matrix}x^m_k &  y^m_k & \dot{x}^m_k &
\dot{y}^m_k\end{matrix}\right]^\intercal
\end{equation}
where $(x^m_k,y^m_k)$ and $(\dot{x}^m_k,\dot{y}^m_k)$ are its
position and velocity in Cartesian coordinates, respectively. The
observer state vector $\xb^o_k$, which is known, is similarly
defined. The dynamic (motion) model is written for the relative
state vector, and is given by:
\begin{equation} \xb_k := \xb^m_k - \xb^o_k =\left[\begin{matrix}x_k &  y_k
& \dot{x}_k & \dot{y}_k\end{matrix}\right]^\intercal.
\label{e:rel_state}
\end{equation}
We adopt a nearly constant velocity (CV) motion model, as a linear
version of (\ref{e:dyn_model}):
\begin{equation}
\xb_{k+1} = \Fb\xb_k -\Ub_{k+1,k} + \Gammab \vb_k \label{e:dyn_eq}
\end{equation}
where
\begin{equation}
\Fb = \left[\begin{matrix}1 & T \\ 0 & 1\end{matrix}\right]\otimes
\Ib_2, \hspace{0.3cm} \Ub_{k+1,k} =
\left[\begin{matrix}x^o_{k+1} - x^o_k - T\dot{x}^o_k \\
y^o_{k+1} - y^o_k - T\dot{y}^o_k \\
\dot{x}^o_{k+1}-\dot{x}^o_k \\
\dot{y}^o_{k+1}-\dot{y}^o_k \end{matrix}\right],
\hspace{0.3cm}\Gammab = \left[\begin{matrix} T^2/2 \\
T\end{matrix}\right]\otimes \Ib_2.
\end{equation}
Explanation: $\otimes$ is the Kroneker product; $\Ib_n$ is identity
matrix of dimension $n$; $T=t_{k+1}-t_{k}$ is the (constant)
sampling interval; $\Fb$ is the transition matrix; $\Ub_{k+1,k}$ is
a known deterministic matrix taking into account the effect of
observer accelerations; $\vb_k \sim {\cal N} (\mathbf{0},\Qb)$ is
white Gaussian process noise with $\Qb=\sigma_v^2\Ib_2$.

The available measurement at time $k$ is the bearing measurement
from the observer platform to the target, referenced clockwise
positive to the $y$-axis and specified by a scalar version of
(\ref{e:meas_model}):
\begin{equation}
z_k = h(\xb_k) + w_k,
\end{equation}
where $h(\xb_k) = \text{atan2}(x_k,y_k)$ is the  four-quadrant
inverse tangent function and $w_k$ is zero-mean white Gaussian noise
with variance $\sigma_w^2$.

\begin{figure}[tbh]
\centerline{\includegraphics[height=7cm]{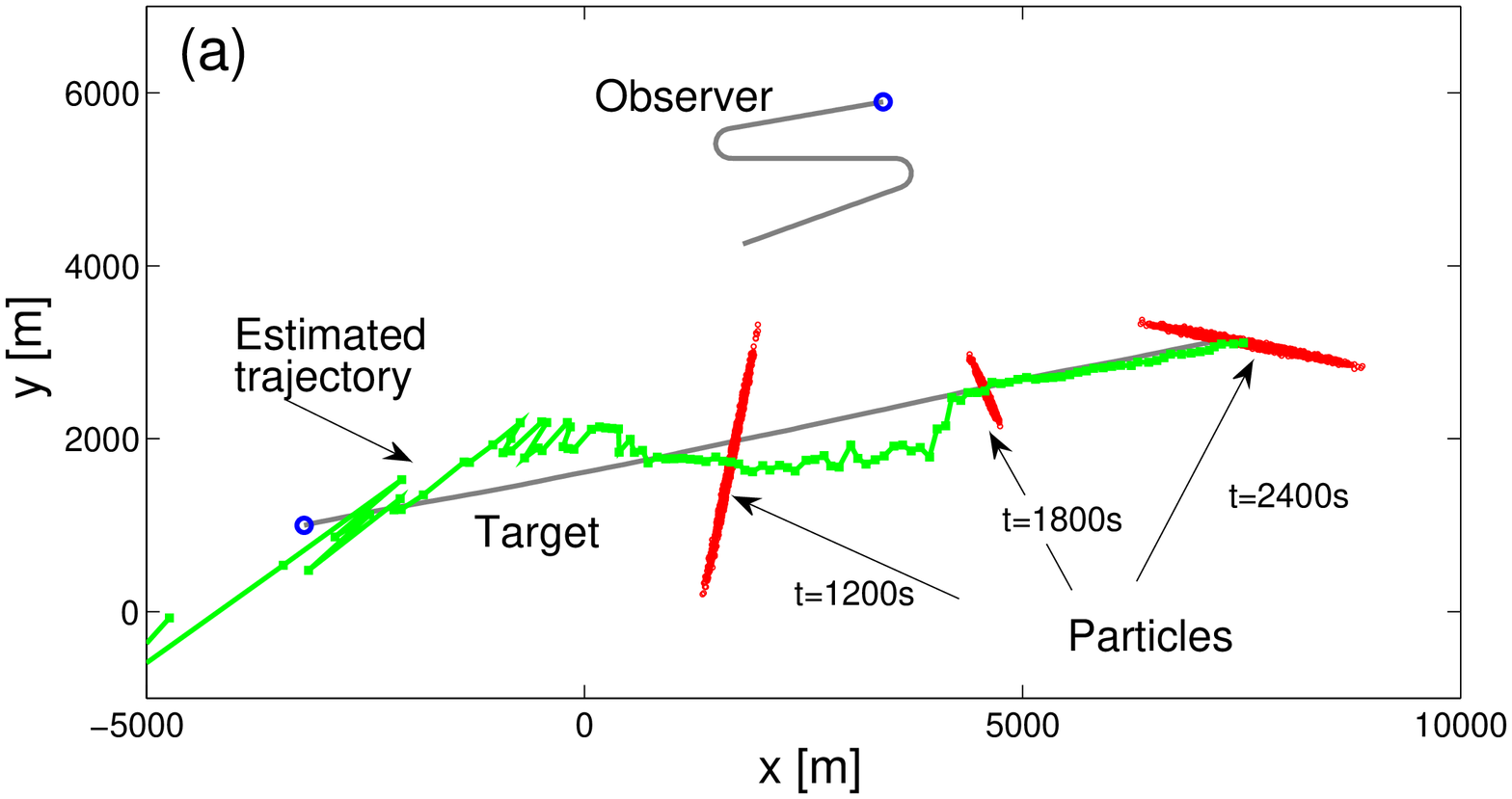}}
\vspace{-0.8cm}\centerline{\includegraphics[height=4.2cm]{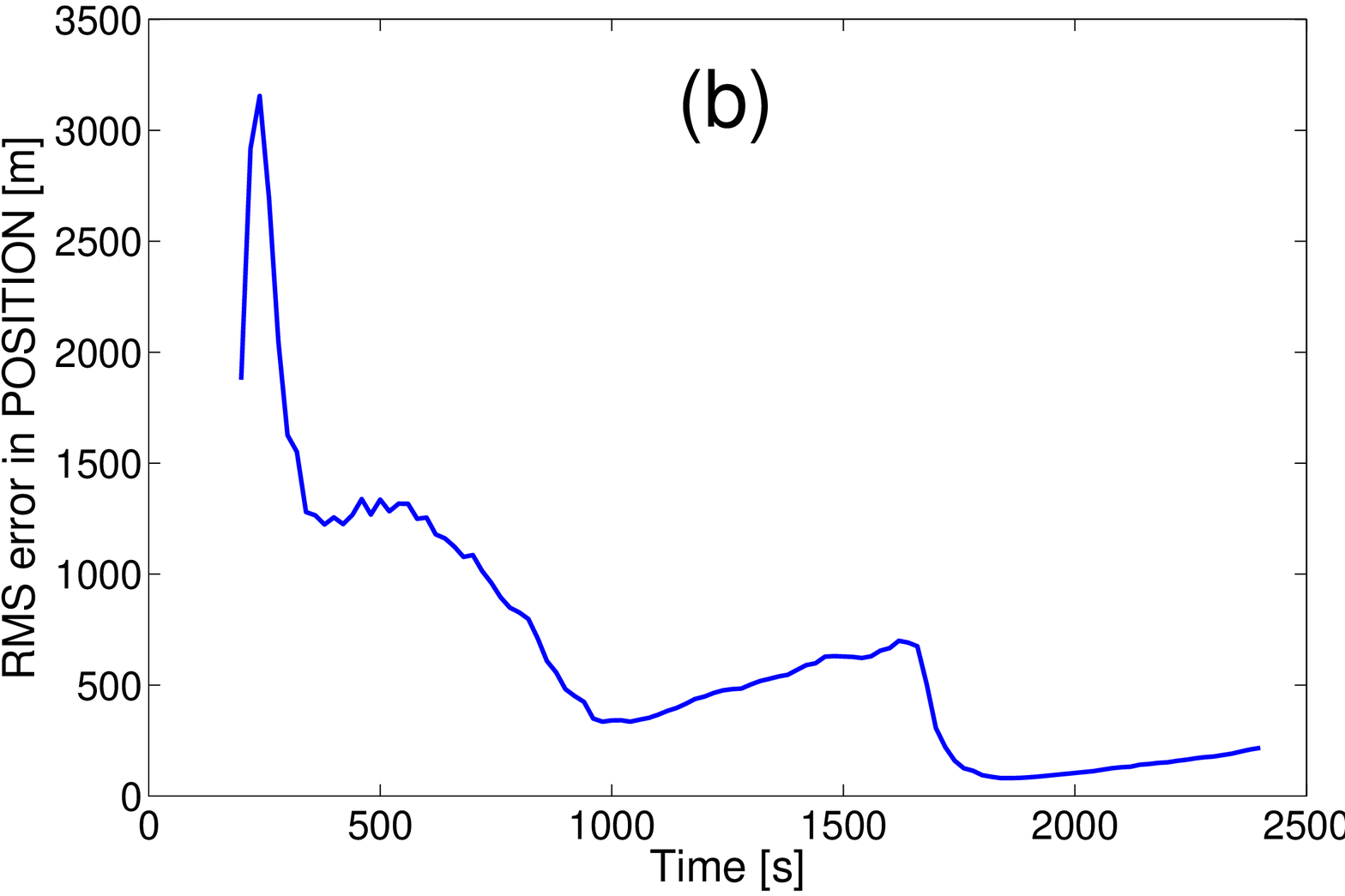}\hspace{2mm}
\includegraphics[height=4.2cm]{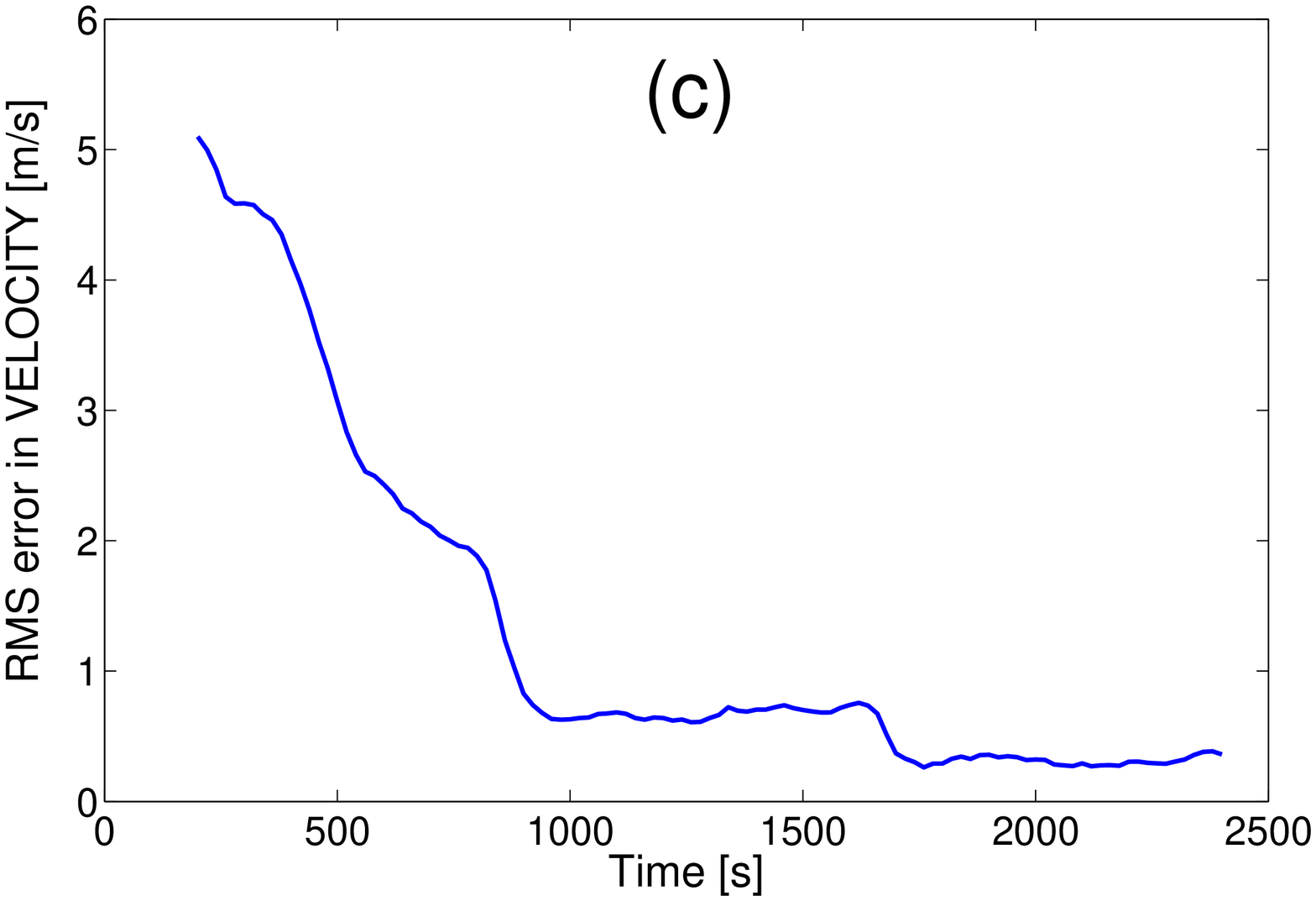}}
 \caption{Particle filter applied to bearings-only filtering problem ($\sigma_w=0.3^o$): (a) top-down view of the scenario and
 the estimated trajectory (red dots are particle clouds); (b) RMS error in position; (c) RMS error in velocity} \label{f:1}
\end{figure}

Fig. \ref{f:1} illustrates the scenario and the performance of the
particle filter. Fig. \ref{f:1}.(a) shows the top-down
observer-target geometry, together with the EAP estimates from the
PF.  This scenario was copied from \cite{beard_15}, but considers
only one of the targets (the full scenario with four targets that appear/disappear at different times, will be used later). The circles indicate the
starting points of the two trajectories. Red dots indicate the cloud
of particles at $t=1200$[s], $t=1800$[s] and $t=2400$[s].
Figs. \ref{f:1}.(b) and (c) display the positional and velocity RMS
error (over time), obtained by averaging over 100 Monte Carlo runs.
The parameters used in simulations were as follows: $\sigma_w =
0.3^o$, $T=20$[s], $\sigma_v=0.005$ [m/s$^2$], $N=5000$ particles.
Nonlinear filtering was carried out during the period when the
target was present, that is from $t=200$[s] to $t=2400$[s], which
corresponds to $111$ discrete-time steps. Note that
Fig. \ref{f:1}.(a) also shows the clouds of particles at time
instances $1200$[s], $1800$[s] and $2400$[s].

The particle filter is initalised by drawing samples from the {\em
initial} or {\it birth} density designed using the first bearing
measurement $z_1$, as well as prior knowledge of (i) the sensor
range $r_{\max}$ and (ii) the maximum speed of the target
$v_{\max}$. The birth density in target position is the sector of a
circle (``pizza slice''), centred at the observer position at $k=1$,
i.e. $(x_1^o,y_1^o)$, determined by the angles $z_1\pm 3\sigma_w$ and
the radius $r_{\max} = 10000$[m]. The target birth density in
velocity is uniform PDF  $\mathcal{U}[-v_{\max},+v_{\max}]$ in both
$\dot{x}^m_1$ and $\dot{y}^m_1$, with $v_{\max}=7.5$ [m/s].

\section{Background: FISST}
\label{s:RFS}

 The prerequisite for the random finite set formulation
of the Bayes-optimal filter is advanced mathematics, referred to as
{\em finite set statistics} (FISST) \cite{mahler_07}, developed by
Ron Mahler. This section will introduce only the bare minimum
necessary to follow the rest of the paper. The full theoretical
details are beyond the scope of this paper, however the proofs
and derivations can be found in Mahler's
excellent books \cite[Part II]{mahler_07} and \cite[Part
I]{mahler_14}.

\subsection{Random finite sets}
\label{s:rfs}
 A random finite set is a convenient probabilistic model
for the representation of multiple stochastic dynamic systems
(objects) and sensor measurements. Suppose that at discrete-time $k$
there are $n_k$ objects with states $\xb_{k,1},\dots,\xb_{k,n_k}$,
taking values in the state space ${\mathcal X}\subseteq
\mathcal{R}^{n_x}$. Both the number of dynamic objects $n_k$ and
their individual states in ${\mathcal X}$ are random and
time-varying. The multi-object state at $k$ is  a finite set
\begin{equation}
\Xb_k   =  \{\xb_{k,1},\dots,\xb_{k,n_k} \} \in {\mathcal
F(\mathcal{X})},
\end{equation}
which can  be modelled as a random finite set on ${\mathcal X}$.
Here ${\mathcal F(\mathcal{X})}$ is the set of finite subsets of
${\mathcal X}$.

Typically, the detection process is imperfect, meaning that not all
of the objects in $\Xb_k$ are detected, while spurious (false)
detections may be created due to background noise and
interferences. Suppose that $\Zb_k$ is a measurement set from such a
detector, reported at time $k$. Then
\begin{equation} \Zb_k = \{\zb_{k,1},\dots,\zb_{k,m_k} \} \in
{\mathcal F(\mathcal{Z})},   \label{e:rfs_meas}
\end{equation}
can also be modelled by a random finite set on the observation space
${\mathcal Z}\subseteq \mathcal{R}^{n_z}$. Both the cardinality
$m_k=|\Zb_k|$ and the individual states in ${\mathcal Z}$ are
random. Here ${\mathcal F(\mathcal{Z})}$ is the set of finite
subsets of ${\mathcal Z}$.

A random finite set (RFS) variable is a random variable that takes
values as unordered finite sets. The cardinality of an RFS variable
$\Xb$ is random and modelled by a discrete distribution $\rho(n) =
Pr\{|\Xb|=n\}$, where $n$ is a non-negative integer,
$n\in\textsf{N}\cup \{0\}$. An RFS $\Xb$ is completely specified by
its cardinality distribution $\rho(n)$ and a family of symmetric
joint distributions\footnote{A joint distribution function
$p_n(\xb_1,\dots,\xb_n)$ is said to be symmetric if its value
remains unchanged for all of the $n!$ possible permutations of its
variables.} $p_n(\xb_1,\dots,\xb_n)$, that characterise the
distribution of its elements over the state space, conditioned on
cardinality $n$.

Since an RFS  is nothing but a finite-set valued random variable, the usual
probabilistic descriptors of a random variable, such as the PDF and
its statistical moments, can be defined for it. FISST provides the
tools for mathematical representation and manipulation of random
finite sets\footnote{Note that while the FISST densities are not
probability densities, they have been shown to be equivalent to
probability densities on ${\mathcal F(X)}$ relative to some
reference measure \cite{vo_05}. Subsequently, we do not distinguish
between FISST densities and probability densities of random finite
sets.}. The FISST PDF of an RFS variable $\Xb$ is denoted $f(\Xb)$
and defined as:
\begin{equation}
 f(\{\xb_1,\dots,\xb_n\}) = n!
\cdot \rho(n)\cdot p_n(\xb_1,\dots,\xb_n).  \label{e:fisst}
\end{equation}
For example, $f(\emptyset) = \rho(0)$, $f(\{\xb\}) = \rho(1)p(\xb)$,
$f(\{\xb_1,\xb_2\}) = 2\rho(2)\,p_2(\xb_1,\xb_2)$, etc. Note that
$f(\emptyset)$,  $f(\{\xb\})$, $f(\{\xb_1,\xb_2\})$, etc, have
different units. Throughout the text, we will refer to the FISST densities  as to the
multi-object PDFs.

Being a PDF, $f(\Xb)$ should integrate to one.  However, the
standard integral cannot be applied; instead we need to introduce
 the {\em set integral}, defined as follows:
\begin{equation}
\int f(\Xb)\, \delta\Xb  :=  f(\emptyset) + \sum_{n=1}^\infty
\frac{1}{n!}\int f(\{\xb_1,\cdots,\xb_n\})d\xb\cdots d\xb_n
\label{e:setintegral} \end{equation} Now it is straightforward to
verify that indeed $f(\Xb)$ integrates to $1$:
\begin{eqnarray} \int f(\Xb)\, \delta\Xb  & = & \rho(0)
+ \sum_{n=1}^\infty \rho(n)\underbrace{\int
p_n(\xb_1,\dots,\xb_n)d\xb_1\cdots d\xb_n}_{= 1,\text{ being a standard PDF}} \nonumber \\
& = &  \sum_{n=0}^\infty \rho(n)  = 1\hspace{.5cm}\text{(since
$\rho(n)$ is a discrete distribution)} \nonumber
\end{eqnarray}

The  cardinality distribution $\rho(n)$ of an RFS variable $\Xb$ can
be obtained from the multi-object PDF $f(\Xb)$ as:
\begin{equation}
\rho(n) = \frac{1}{n!}\int f(\{\xb_1,\dots,\xb_n\}) d\xb_1,\dots
d\xb_n
\end{equation}

The {\em intensity function} (also known as the probability
hypothesis density or PHD)  is an important characterisation of  an
RFS $\Xb$ on $\mathcal{X}$,  defined as its first-order statistical
moment.
In order to define the PHD function, let us first introduce the set Dirac delta function
\[\delta_{\Xb}(\xb)=\sum_{\wb\in\Xb}\delta_{\wb}(\xb)\]
with $\delta_{\wb}(\xb)$ being the standard Dirac delta function concentrated at $\wb$.
Now we can  express the cardinality of an RFS $\Xb$ as follows:
\begin{equation}
|\Xb| = \displaystyle{\int_{\mathcal{X}}} \delta_{\Xb}\!\left( \xb \right) d \xb \,.
\end{equation}
We would like to define the PHD function $D\!\left( \xb \right)$ of $\Xb$ in such a way that the expected cardinality of $\Xb$ over the state space $\mathcal{X}$ is obtained as the integral
\begin{equation}
\operatorname{\mathcal{E}}\!\left\{|\Xb|  \right\} = \displaystyle \int_{\mathcal{X}} D\!\left( \xb \right) d \xb \, .
\label{eq:objnum}
\end{equation}
Since
\begin{eqnarray}
\operatorname{\mathcal{E}}\!\left\{ |\Xb| \right\} & \triangleq & \displaystyle \int |\Xb| f\!\left( \Xb \right) \delta \Xb\\
			& = & \displaystyle \int \left[ \int_{\mathcal{X}} \delta_{\Xb}\!\left( \xb \right) d\xb \right] f\!\left( \Xb \right) \delta \Xb \\
			& = & \displaystyle \int_{\mathcal{X}} \left[ \int \delta_{\Xb}\!\left( \xb \right) f\!\left( \Xb \right) \delta \Xb \right] d \xb \, ,
	\label{eq:d1}
\end{eqnarray}
comparing (\ref{eq:objnum}) with (\ref{eq:d1}), notice that
\begin{equation}
D\!\left( \xb \right) \triangleq \operatorname{\mathcal{E}}\!\left\{ \delta_{\Xb}\!\left( \xb \right) \right\} = \int \delta_{\Xb}\!\left( \xb \right) f\!\left( \Xb \right) \delta \Xb.
	\label{eq:def-d}
\end{equation}
Here $\operatorname{\mathcal{E}}\!\left\{ \delta_{\Xb}\!\left( \xb \right) \right\}$ is the expected value (the first statistical moment) of the RFS $\Xb$. Note that $D(\xb)$ is a density function on the state space $\mathcal{X}$. 	

In general, the PHD function $D\!\left( \xb \right)$ and the cardinality distribution $\rho\!\left( n \right)$ do not completely characterize the multi-object PDF. However, we will see in the next section that for some specific RFSs, the characterization is complete.


\subsection{Some common RFS variables}
\label{s:common}

\paragraph{\bf Bernoulli RFS}  This RFS can either be empty (with probability $1-r$) or have
one element (with probability $r$), spatially distributed over
$\mathcal{X}$ according to the (standard) PDF $p(\xb)$. The FISST
PDF of the Bernoulli RFS $\Xb$ is given by:
\begin{equation}
f(\Xb) = \begin{cases} 1-r, & \text{ if } \Xb=\emptyset\\
r\cdot p(\xb), & \text{ if } \Xb = \{\xb\}. \end{cases}
\label{e:Bern_PDF}
\end{equation}
The intensity function of the Bernoulli RFS $\Xb$ is $D(\xb) =
r\cdot p(\xb)$.

\paragraph{\bf IID Cluster RFS}
Let the cardinality distribution  of independent identically
distributed (IID) cluster RFS $\Xb$ be $\rho(n)$. For a given
cardinality, the elements of $\Xb$ are each IID random variables
distributed according to the (standard) PDF $p(\xb)$ on
$\mathcal{X}$. The multi-object PDF of $\Xb$ is:
\begin{equation}
f(\Xb) = |\Xb|!\cdot \rho(|\Xb|)\, \prod_{\xb\in\Xb}p(\xb)
\label{e:iid_cluster_pdf}
\end{equation}
The main simplification in comparison with (\ref{e:fisst}) is that
due to the IID property, the symmetric joint distribution is
replaced by a product. The intensity function of an IID cluster RFS
is: \begin{equation}
 D(\xb) = p(\xb)\sum_{n=1}^\infty n \, \rho(n)
 \label{e:int_iid_cl}
\end{equation}

\paragraph{\bf Poisson RFS} If the cardinality distribution of an IID cluster
RFS is Poisson with parameter $\lambda>0$, i.e. \begin{equation}
\rho(n) = \frac{e^{-\lambda}\lambda^n}{n!}, \hspace{1cm}
n=0,1,2,\dots \label{e:rho_poisson}\end{equation}
 then this RFS is referred to as the
Poisson RFS $\Xb$. Its multi-object PDF follows from
(\ref{e:iid_cluster_pdf}) and is given by:
\begin{equation} f(\Xb) = e^{-\lambda}\prod_{\xb\in\Xb}\lambda\,
p(\xb) \label{e:poisson_proc_PDF}
\end{equation}
while its intensity function is
\begin{equation} D(\xb) = \lambda\,
p(\xb). \label{e:D_poiss}\end{equation} The Poisson RFS is the only
RFS which is completely specified by its intensity function, because
knowledge of $\lambda$ and $p(\xb)$ fully determines the multi-object PDF
in (\ref{e:poisson_proc_PDF}). Both $\lambda$ and $p(\xb)$ can be worked out from
$D(\xb)$: $\lambda$ is the expected cardinality of $\Xb$, i.e.
$\mathcal{E}\{|\Xb|\} = \int D(\xb)d\xb = \lambda$, while  $p(\xb) =
D(\xb)/\lambda$.


\paragraph{\bf Multi-Bernoulli RFS}  A multi-Bernoulli RFS $\Xb$ is a
union of independent Bernoulli RFSs $\Xb_i$,  each characterized by
existence probability  $r_i$ and the spatial PDF $p_i(\xb)$, for
$i=1,\dots,M$:
\[ \Xb = \bigcup_{i=1}^M \Xb_i. \]
The multi-object PDF of $\Xb=\{\xb_1,\dots,\xb_n\}$ is given by:
\begin{equation}
f(\Xb) = n! \,\left[\prod_{j=1}^M(1-r_j)\right]\cdot\sum_{1\leq i_1<
\cdots < i_{n}\leq M}\;\frac{r_{i_1}\,
p_{i_1}(\xb_1)}{1-r_{i_1}}\cdots \frac{r_{i_n}\,
p_{i_n}(\xb_n)}{1-r_{i_n}}. \label{e:mbpdf}
\end{equation}
For example, $f(\emptyset) = \prod_{j=1}^M (1-r_j)$, $f(\{\xb\}) =
f(\emptyset)\sum_{i=1}^M \frac{r_i\,p_i(\xb)}{1-r_i}$, etc.  Note
that the pairs $\big(r_i,p_i(\xb)\big)$, $i=1,\dots,M$, fully
specify the multi-object PDF (\ref{e:mbpdf}).  Consequently, they also
determine the intensity function of the multi-Bernoulli RFS, which
is given by: $D(\xb)=\sum_{i=1}^M r_i\, p_i(\xb)$.

\section{Particle methods for the RFS Bayes-optimal filter}
\label{s:4}

\subsection{Formulation of the RFS Bayes-optimal filter}

The goal of  the RFS Bayes-optimal filter is to estimate the
posterior density of a multi-object state, represented by the RFS
variable $\Xb_k$. Evolution of $\Xb_k$ is modelled by a Markov
process, characterized by its initial FISST density $f_0(\Xb_0)$ and
the FISST transitional density $\Pi_{k|k-1}(\Xb_k|\Xb_{k-1})$. The
standard case of $\Pi_{k|k-1}(\Xb_k|\Xb_{k-1})$ has been derived in
Chapter 13 of \cite{mahler_07} as the union of a
multi-Bernoulli object survival RFS and a Poisson birth RFS, under
the assumption that the constituent RFSs are mutually independent.
It can be expressed in a compact form as:
\begin{equation}
\Pi_{k|k-1}(\Xb|\Xb') = f_b(\Xb)(1-p_\sS)^{|\Xb'|} \sum_{\theta}
\prod_{i:\theta(i)>0} \frac{p_{\sS}\cdot
\pi_{k|k-1}(\xb_{\theta(i)}|\xb'_i)}{(1-p_{\sS})\cdot \mu_0\,
b_k(\xb_{\theta(i)})}  \label{e:mo_trans}
\end{equation}
where $\theta:\{1,\dots,|\Xb'|\}\rightarrow\{0,1,\dots,|\Xb|\}$
represents a list of all possible assignments of elements from the set
$\Xb'$ to the elements of the set $\Xb$ (here the assumptions are that
$|\Xb'|>0$ and $|\Xb|\geq 0$); $p_{\sS}$ is the probability of
object survival from time $k-1$ to $k$; $f_b(\Xb)$ is the object
birth multi-object PDF, which assuming a Poisson birth process with the
mean rate $\mu_0$ and distribution $b_k(\xb)$, according to
(\ref{e:poisson_proc_PDF}) is given by:
\[ f_b(\Xb) = e^{-\mu_0}\prod_{\xb\in\Xb}\mu_0\, b_k(\xb). \]
Finally, $\pi_{k|k-1}(\xb|\xb')$ is the (standard) transitional
density, defined in Sec. \ref{s:2}.

The multi-object state $\Xb_k$ is not observed directly, but through
the observation process, assumed to be conditionally independent
given the multi-object state process, and fully specified by the (multi-object)
likelihood function. Many different observation models have been
considered in the literature, such as the models for intensity
measurements versus detector-output measurements\footnote{An intensity measurement at time $k$ represents a raw sensor measurement, prior to detection thresholding. In tracking literature this is also referred to as track-before-detect approach. Using intensity measurements (rather than the detector-output measurements) can lead to the better error performance, but at increased computational load due to the large data flow.  An intensity measurement at time $k$ is a vector, rather than a random finite set.  A detailed explanation with examples of intensity measurement models can be found in \cite{bf_tute}.}, extended versus point
target measurements, finite resolution measurements, fuzzy or
imprecise measurements, see for details \cite{bf_tute},
\cite{springer_book_13}, \cite{granstrom2012extended}, \cite{mahler_14}, \cite{beard_15}.

The detector-output  measurement model for a point-size target,
referred to as the standard model \cite{mahler_07},\cite{mahler_14},
will be adopted throughout this tutorial. Recall from
(\ref{e:rfs_meas}) that the standard measurement is modelled by an
RFS variable $\Zb_k$. The
 likelihood function $\varphi_k(\Zb_k|\Xb_k)$ for this model has been derived in Chapter 12 of
 \cite{mahler_07} as the superposition of a multi-Bernoulli object-detection RFS and a Poisson clutter
 (false detection) RFS. Under the assumption that the constituent RFSs are mutually independent, the multi-object  likelihood function
can be expressed by:
\begin{equation}
\varphi_{k}(\Zb|\Xb) = f_c(\Zb)(1-p_\sD)^{|\Xb|} \sum_{\theta}
\prod_{i:\theta(i)>0} \frac{p_{\sD}\cdot
g_{k}(\zb_{\theta(i)}|\xb_i)}{(1-p_{\sD})\cdot \lambda\,
c(\zb_{\theta(i)})}.  \label{e:mo_lik}
\end{equation}
Explanation:
$\theta:\{1,\dots,|\Xb|\}\rightarrow\{0,1,\dots,|\Zb|\}$ represents
a list of all associations of elements from set $\Xb$ to the
elements of set $\Zb$ (here the assumption is that if $\xb_i\in\Xb$
is not detected, then $\theta(i)=0$; also, a target $\xb\in\Xb$ can generate at most one measurement $\zb\in\Zb$); $p_{\sD}$ is the probability
of detection; $f_c(\Zb)$ is the  multi-object PDF of clutter, which is
assumed to be a Poisson RFS with the mean rate $\lambda$ and
distribution $c(\zb)$ over $\mathcal{Z}$. According to
(\ref{e:poisson_proc_PDF}), the clutter multi-object PDF is:
\begin{equation}
 f_c(\Zb) = e^{-\lambda}\prod_{\zb\in\Zb}\lambda\,
c(\zb). \label{e:clutter_PDF} \end{equation} Finally,
$g_{k}(\zb|\xb)$ in (\ref{e:mo_lik}) is the conventional likelihood
function, introduced in Sec. \ref{s:2}.

%

Given $\Xb_k$, measurement
$\Zb_k$ is assumed to be statistically independent of
$\Zb_\ell$, where $\ell\neq k$.
The Bayes-optimal filtering problem can now be cast in the random
finite set framework. Suppose that at time $k-1$ the posterior FISST
PDF of the multi-object state, $f_{k-1|k-1}(\Xb_{k-1}|\Zb_{1:k-1})$
is known. Here $\Zb_{1:k-1}\equiv \Zb_1,\dots,\Zb_{k-1}$ is the
sequence of all previous measurements. Then the predicted and
updated posterior multi-object densities  can be expressed as
follows ~\cite{mahler_07}:
\begin{eqnarray}
f_{k|k-1}(\Xb_{k}|\Zb_{1:k-1})  & =  & \int
\Pi_{k|k-1}(\Xb_{k}|\Xb') f_{k-1|k-1}(\Xb'|\Zb_{1:k-1})\delta \Xb' \label{e:pred_PDF}\\
f_{k|k}(\Xb_{k}|\Zb_{1:k}) & =  &
{\frac{\varphi_k(\Zb_{k}|\Xb_{k}) f_{k|k-1}(\Xb_{k}|\Zb_{1:k-1})
}{ \int \varphi_k(\Zb_{k}|\Xb) f_{k|k-1}(\Xb|\Zb_{1:k-1})\delta
\Xb}}, \label{e:upd_PDF}
\end{eqnarray}
respectively.

The recursion (\ref{e:pred_PDF})-(\ref{e:upd_PDF})  is a non-trivial
generalisation of (\ref{e:pred_eq})-(\ref{e:upd_eq}), because
the integrals in (\ref{e:pred_PDF})-(\ref{e:upd_PDF})  are set
integrals and the expressions for $\Pi_{k|k-1}(\Xb_k|\Xb_{k-1})$ and
$\varphi_k(\Zb_k|\Xb_k)$ are quite involved. Computing the exact
multi-object posterior density is numerically intractable and all
practical algorithms are based on approximations. An important
feature of the RFS Bayes-optimal filter is that it performs
multi-object filtering, as opposed to target (object) tracking. The
difference is significant: a point estimate from the multi-object
filter, at each time step, is a collection of (unlabelled and
unordered) object state estimates; a multi-target tracker, on the
other hand, produces labelled state trajectory estimates, or tracks.

\subsection{Particle method approximations}

Implementation of the RFS Bayes-optimal filter using the particle
method has been considered in the past, both assuming the intensity
measurement model (typically without a reference to RFS models, e.g.
\cite{boers_04}) and the standard measurement model
\cite{sidenbladh_03}, \cite{zajic_03}, \cite{vo_05},
\cite{vihola_07}, \cite{automatica_10}, \cite{reuter_11}. Since the
filter is defined on the set of subsets ${\mathcal F(\mathcal{X})}$,
it is computationally very demanding and practical only for a small
number of objects. A particle in the state space ${\mathcal
F(\mathcal{X})}$ can be expressed as:
\begin{equation}
\Xb_k^{(i)} = \begin{cases} \emptyset, & \text{if }
|\Xb_k^{(i)}|=0\\
                               [\xb_k^{(i)}], & \text{if } |\Xb_k^{(i)}|=1\\
                                [(\xb_{k,1}^{(i)})^\intercal\;\; (\xb_{k,2}^{(i)})^\intercal]^\intercal, & \text{if } |\Xb_k^{(i)}|=2\\
                               \dots & \\
                               [(\xb_{k,1}^{(i)})^\intercal \;\; (\xb_{k,2}^{(i)})^\intercal \;\; \dots \;\;
(\xb^{(i)}_{k,\nu_{\max}})^\intercal]^\intercal, & \text{if }
|\Xb_k^{(i)}|=\nu_{\max}
\end{cases}
\label{e:stacked}
\end{equation}
where $\nu_{\max}$ denotes the maximum number of targets (a design
parameter). Note that particle $i=1,2,\dots,N$, for a given
cardinality, is represented in (\ref{e:stacked}) by a vector. While
this was done to simplify computer implementation, one should keep
in mind that the multi-object particle essentially represents a
random finite set, that is, any permutation of its elements
(objects, targets) results in the equivalent particle. By ignoring
the permutation equivalence, the multi-object state particle filter
can be affected by the {\em mixed labelling} problem, see
\cite{boers_mixed_labeling}. It has been noted that mixed labeling
is typically resolved after few time steps due to resampling step in
the particle filter \cite{boers_mixed_labeling}.

The pseudo-code of the bootstrap-type particle filter which
implements a cycle of the RFS Bayes-optimal filter at time $k$ is
given in Alg. \ref{a:mopf}. The proposed multi-object particles at
time $k$ are constructed based on the transitional density, the
survival probability and the birth distribution of new targets,
ignoring the association events in (\ref{e:mo_trans}). The average
number of newborn targets is $\mu_0$, with birth distribution
$b_k(\xb)$ in line 14 typically designed using the measurements from
time $k-1$. The computation of the multi-object likelihood
$\varphi_{k}(\Zb_k|{\Xb}^{(i)}_{k})$ in line 17, defined in
(\ref{e:mo_lik}),  requires the evaluation all association hypotheses
$\theta$ between objects in ${\Xb}^{(i)}_{k}$ and detections
in $\Zb_k$. The number of these hypotheses grows exponentially with
the number of targets. For example, if the cardinalities are:
$|{\Xb}^{(i)}_{k}|=2$ and $|\Zb_k|=3$, the number of
association hypothesis \footnote{In this case $\theta: \{\xb_1,\xb_2\}\rightarrow \{\emptyset, \zb_1, \zb_2,\zb_3\}$, because targets can be undetected. The list of possible assignments is: $\theta_1: \xb_1\rightarrow \emptyset; \xb_2\rightarrow \emptyset$; $\theta_2: \xb_1\rightarrow \emptyset; \xb_2\rightarrow \zb_1$;  $\theta_3: \xb_1\rightarrow \emptyset; \xb_2\rightarrow \zb_2$;  $\theta_4: \xb_1\rightarrow \emptyset; \xb_2\rightarrow \zb_3$;  $\theta_5: \xb_1\rightarrow \zb_1; \xb_2\rightarrow \emptyset$; $\theta_6: \xb_1\rightarrow \zb_1; \xb_2\rightarrow \zb_2$; $\theta_7: \xb_1\rightarrow \zb_1; \xb_2\rightarrow \zb_3$; $\theta_8: \xb_1\rightarrow \zb_2; \xb_2\rightarrow \emptyset$; $\theta_9: \xb_1\rightarrow \zb_2; \xb_2\rightarrow \zb_1$; $\theta_{10}: \xb_1\rightarrow \zb_2; \xb_2\rightarrow \zb_3$; $\theta_{11}: \xb_1\rightarrow \zb_3; \xb_2\rightarrow \emptyset$; $\theta_{12}: \xb_1\rightarrow \zb_3; \xb_2\rightarrow \zb_1$; $\theta_{13}: \xb_1\rightarrow \zb_3; \xb_2\rightarrow \zb_2$.} is $13$. This exponentially growing
computational cost is the major limitation of the RFS Bayes-optimal
filter in practice.

\begin{algorithm}[ht]
\caption{ Pseudo-code of the RFS bootstrap filter} {\footnotesize
\begin{algorithmic}[1]
\Function{RFS bootstrap filter}{} \State Input:
$\{\Xb^{(i)}_{k-1}\}_{1 \leq i \leq N}$,  $\Zb_k$ \For
{$i=1,\dots,N$}
    \State $\tilde{\Xb}^{(i)}_{k} = [\;]$
    \For {$j=1,\dots,|\Xb^{(i)}_{k-1}|$}
        \State $u_*\sim\mathcal{U}_{[0,1]}$
        \If {$u_* < p_{\sS}$}
             \State $\xb_* \sim \pi_{k|k-1}(\xb|\xb^{(i)}_{k-1,j})$
             \State $\tilde{\Xb}^{(i)}_{k} =
                      \left[\left(\tilde{\Xb}^{(i)}_{k}\right)^\intercal \;\;
                       \xb_*^\intercal\right]^\intercal$
        \EndIf
    \EndFor
    \State $\mu \sim \text{Poisson}(\mu_0)$
    \For {$j=1,\dots,\mu$}
            \State $\xb_* \sim b_k(\xb)$
             \State $\tilde{\Xb}^{(i)}_{k} =
                      \left[\left(\tilde{\Xb}^{(i)}_{k}\right)^\intercal \;\;
                       \xb_*^\intercal\right]^\intercal$
    \EndFor
    \State  Calculate weight: $\tilde{w}^{(i)}_k = \varphi_{k}(\Zb_k|\tilde{\Xb}^{(i)}_{k})$
\EndFor \State $w_k^{(i)} = \tilde{w}_k^{(i)}/\sum_{j=1}^N
\tilde{w}_k^{(j)}$, for  $i=1,\dots,N$
 \For {$i=1,\dots,N$}
 \State Select index $j^i\in\{1,\dots,N\}$ with
probability $w_k^{(i)}$ \State
 $\Xb^{(i)}_k = \tilde{\Xb}^{(j_i)}_{k}$
 \EndFor
\State Output:  $\{\Xb^{(i)}_{k}\}_{1\leq i \leq N}$ \EndFunction
\end{algorithmic}
} \label{a:mopf}
\end{algorithm}

Estimation of the multi-object state $\Xb_k$ from the particle
approximation of the posterior $f_{k|k}(\Xb_{k}|\Zb_{1:k})$ is
discussed briefly next. The cardinality distribution
 can be estimated as:
\begin{equation}
\hat{\rho}_{k|k}(n|\Zb_{1:k}) =
\frac{1}{N}\sum_{i=1}^N \delta\left[|\Xb^{(i)}_k|,n\right],
\hspace{1cm}n=0,1,2,\dots,\nu_{\max},
\label{e:KDF}
\end{equation}
where $\delta[\ell,n]$ is the Kronecker delta function which equals
$1$ if $\ell=n$ and zero otherwise. Practically,  for each
$n=0,1,\dots,\nu_{\max}$, it is necessary to count the number of
multi-target particles whose cardinality equals $n$, and then to
divide this count with the total number of particles $N$. Then one
can estimate the number of targets in $\Xb_k$, that is
$\hat{n}_{k|k}$, either as the expected value or the maximum of the
posterior $\hat{\rho}_{k|k}(n|\Zb_{1:k})$. Given $\hat{n}_{k|k}$,
the multi-target state can be estimated from the particle system
$\{w_k^{(i)},\Xb_k^{(i)}\}_{1 \leq i \leq N}$ as
\begin{equation}
\hat{\Xb}_k = \frac{\sum_{i=1}^N w^{(i)}_k\Xb^{(i)}_k\,
\delta\left[|\Xb^{(i)}_k|,\hat{n}_k\right]}{\sum_{i=1}^N w^{(i)}_k\,
\delta\left[|\Xb^{(i)}_k|,\hat{n}_k\right]},
\end{equation}
which represents the mean of the multi-target particles
characterized by cardinality $\hat{n}_{k|k}$.

\subsection{Bernoulli particle filter}
\label{s:bpf}
The Bernoulli filter is the special case of the  RFS
Bayes-optimal filter, derived by Mahler \cite{mahler_07} under the
assumption that $\Xb_k$ is a Bernoulli RFS. Recall that the
Bernoulli RFS $\Xb$ can have zero or one element and its PDF
$f(\Xb)$ is completely specified by: (1) the probability of object
existence $r$ and (2) the  PDF $p(\xb)$ on $\mathcal{X}$. Hence, the
Bernoulli filter is the Bayes-optimal filter for joint detection and
tracking of a single object, where detection is carried out by
monitoring the posterior probability of object existence.

A detailed tutorial on Bernoulli filters, their formulation for
different measurement models, their numerical implementation and
various applications can be found in \cite{bf_tute}. Here we present
only the prediction and update equations for the standard
measurement model. Suppose the posterior
$f_{k-1|k-1}(\Xb|\Zb_{1:k-1})$ is known and specified by the pair
$\big(r_{k-1|k-1},p_{k-1|k-1}(\xb)\big)$. The prediction equations
of the Bernoulli filter are given by:
\begin{eqnarray}
r_{k|k-1} & = & p_b(1-r_{k-1|k-1}) + p_{\sS} r_{k-1|k-1} \label{e:BPF1}\\
p_{k|k-1}(\xb) & = &
\frac{p_b(1-r_{k-1|k-1})b_{k|k-1}(\xb)}{r_{k|k-1}} + \nonumber \\
& & \; \; \frac{p_{\sS}r_{k-1|k-1}\int
\pi_{k|k-1}(\xb|\xb')p_{k-1|k-1}(\xb')d\xb'}{r_{k|k-1}}
\label{e:BPF2}
\end{eqnarray}
where $p_b$ is the probability of object ``birth'' from the previous time $k-1$
to $k$ and $b_{k|k-1}(\xb)$ is the object birth density. The update
equations are as follows:
\begin{eqnarray}
r_{k|k} & = & \frac{1-\Delta_k}{1-r_{k|k-1}\Delta_k}\, r_{k|k-1} \label{e:BPF3}\\
p_{k|k}(\xb) & = & \frac{1-p_{\sD} +
p_{\sD}\sum_{\zb\in\Zb_k}\frac{g_k(\zb|\xb)}{\lambda
c(\zb)}}{1-\Delta_k} \; p_{k|k-1}(\xb)
\end{eqnarray}
where, assuming $p_{\sD}$ is independent of the state,
\begin{equation}
\Delta_k = p_{\sD}\left(1 - \sum_{\zb\in\Zb_k}\frac{\int
g_k(\zb|\xb)p_{k|k-1}(\xb)d\xb}{\lambda c(\zb)}\right).
\label{e:Delta}
\end{equation}
It can be verified that if $r_{k-1|k-1}=1$, $p_b=0$, $p_{\sD}=1$ and
$\lambda = 0$, the Bernoulli filter equations reduce to the standard
Bayes filter equations (\ref{e:pred_eq})-(\ref{e:upd_eq}).

The Bernoulli particle filter propagates through time the scalar
value of $r_{k|k}$ and the weighted particle system
$\{w_k^{(i)},\xb_k^{(i)}\}_{1\leq i \leq N}$. The pseudo-code of the
Bernoulli bootstrap-type particle filter is given in Alg. \ref{a:3}.
It implements the birth density  in (\ref{e:BPF2}) as
$b_{k|k-1}(\xb) = \int \pi_{k|k-1}(\xb|\xb')b_{k-1}(\xb')d\xb'$,
where (see line 3 in Alg. \ref{a:3}) the birth density at $k-1$, i.e.
$b_{k-1}(\xb')$ is designed using the measurements in $\Zb_{k-1}$.
 This is done by
drawing $N_m$ particles for each $\zb\in\Zb_{k-1}$, hence $B_{k-1} =
|\Zb_{k-1}|\cdot N_m$. The weights of these ``new-born'' particles
are set to $1/B_{k-1}$. The number of particles from step 5 onwards
is $N+B_{k-1}$, and reduces back to $N$ only by the resampling step
(lines 13-16).

\begin{algorithm}[ht]
\caption{ Pseudo-code of a Bernoulli bootstrap-type particle filter}
{\footnotesize
\begin{algorithmic}[1]
\Function{Bernoulli Particle Filter}{} \State Input: $r_{k-1|k-1}$,
$\{\xb^{(i)}_{k-1}\}_{1\leq i \leq N}$, $\Zb_k$, $\Zb_{k-1}$
   \State Draw: $\xb^{(i)}_{k-1} \sim
b_{k-1}(\xb)$ designed based on $\Zb_{k-1}$ for
$i=N+1,\dots,N+B_{k-1}$ \State Compute the predicted existence
$r_{k|k-1}$ using (\ref{e:BPF1})
   \State Draw a sample: $\xb^{(i)}_{k|k-1} \sim
\pi_{k|k-1}(\xb_k|\xb^{(i)}_{k-1})$  for $i=1,\dots,N+B_{k-1}$
\State Weights: $w^{(i)}_{k|k-1} = \frac{p_{\sS}\, r_{k-1|k-1}}{N\,
r_{k|k-1}}$
   for $i=1,\dots,N$ \Comment 2nd term in (\ref{e:BPF2})
\State Weights: $w^{(i)}_{k|k-1} = \frac{p_b
(1-r_{k-1|k-1})}{B_{k-1}\, r_{k|k-1}}$
   for $i=N+1,\dots,N+B_{k-1}$ \Comment 1st term in (\ref{e:BPF2})
\State For every $\zb\in\Zb_k$ compute $I_k(\zb)=\sum_{i = 1}^{
N+B_{k-1}}w^{(i)}_{k|k-1}\cdot g_k(\zb|\xb^{(i)}_{k|k-1})$ \State
Compute $\delta_k\approx p_{\sD}\left(1-\sum_{\zb\in\Zb_k}
\frac{I_k(\zb)}{\lambda\, c(\zb)}\right)$ \Comment
Eq.(\ref{e:Delta}) \State Compute the updated existence  $r_{k|k}$
using (\ref{e:BPF3})  \State Weights: $\tilde{w}^{(i)}_{k|k} \approx
\left[1-p_{\sD} +
p_{\sD}\sum\limits_{\zb\in\Zb_k}\frac{g_k(\zb|\xb^{(i)}_{k|k-1})}{\lambda\,
c(\zb)}\right]\cdot w^{(i)}_{k|k-1}$, for $i=1,\dots,N+B_{k-1}$
\State Normalise weights: $w^{(i)}_{k|k} =
\frac{\tilde{w}^{(i)}_{k|k}}{\sum_{j=1}^{N+B_{k-1}}\tilde{w}^{(j)}_{k|k}}$,
for $i=1,\dots,N+B_{k-1}$
 \For {$i=1,\dots,N$} \Comment
(Resampling) \State Select index $j^i\in\{1,\dots,N+B_{k-1}\}$ with
probability $w_{k|k}^{(i)}$ \State
 $\xb^{(i)}_k = \xb^{(j_i)}_{k|k-1}$ \EndFor
\State Apply MCMC move to particles and output: $r_{k|k}$,
$\{\xb^{(i)}_{k}\}_{1\leq i \leq N}$ \EndFunction
\end{algorithmic}
} \label{a:3}
\end{algorithm}

\subsection{Demonstration: Bearings-only detection and tracking}
\label{s:bot2}

Let us demonstrate the Bernoulli particle filter in action. The
observer-target scenario is the same as described in
Sec. \ref{s:bot1}, however, this time we consider the full
observation interval from $t=0$ to $t=3000$[s]. Recall that the
target exists only for $200\text{[s]} \leq t \leq 2400\text{[s]}$.
Furthermore, detection probability is $p_{\sD}=0.95$ and the average
number of false bearings-only measurements is $\lambda=1$ per scan.
Clutter distribution is uniform, that is
$c(z)=(2\pi)^{-1}$[rad$^{-1}$]. The parameters of the Bernoulli PF
were set to: $N_m=2500$, $p_b=0.01$, $p_{\sS}=0.98$. The remaining
parameters were the same as in Sec. \ref{s:bot1}: $\sigma_w = 0.3^o$,
$T=20$[s], $\sigma_v=0.005$ [m/s$^2$], $r_{\max}=10000$[m], $N=5000$
particles.

Birth density $b_{k-1}(\xb)$ is designed adaptively using
$\Zb_{k-1}$, to form a mixture density:
\begin{equation}
b_{k-1}(\xb|\Zb_{k-1}) =
\frac{1}{|\Zb_{k-1}|}\sum_{\zb\in\Zb_{k-1}}\beta(\xb|\zb)
\label{e:birth_den}
\end{equation}
Here $\beta(\xb|\zb)$ is the birth density created using the
standard technique for particle filter initialisation when tracking
with bearings-only measurements, explained already in Sec. \ref{s:bot1}.

\begin{figure}[tbh]
\centerline{\includegraphics[height=5cm]{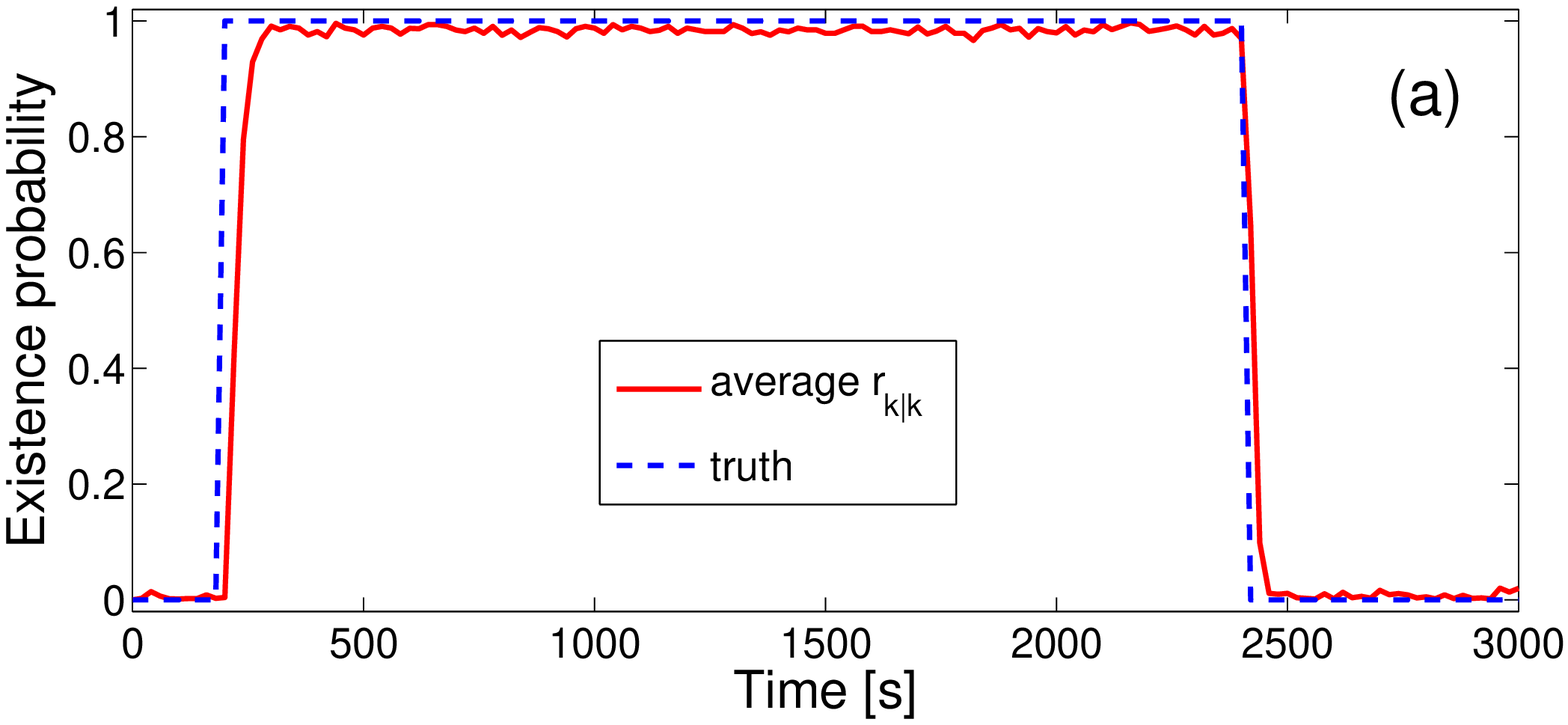}}
\vspace{-0.8cm}\centerline{\includegraphics[height=4.4cm]{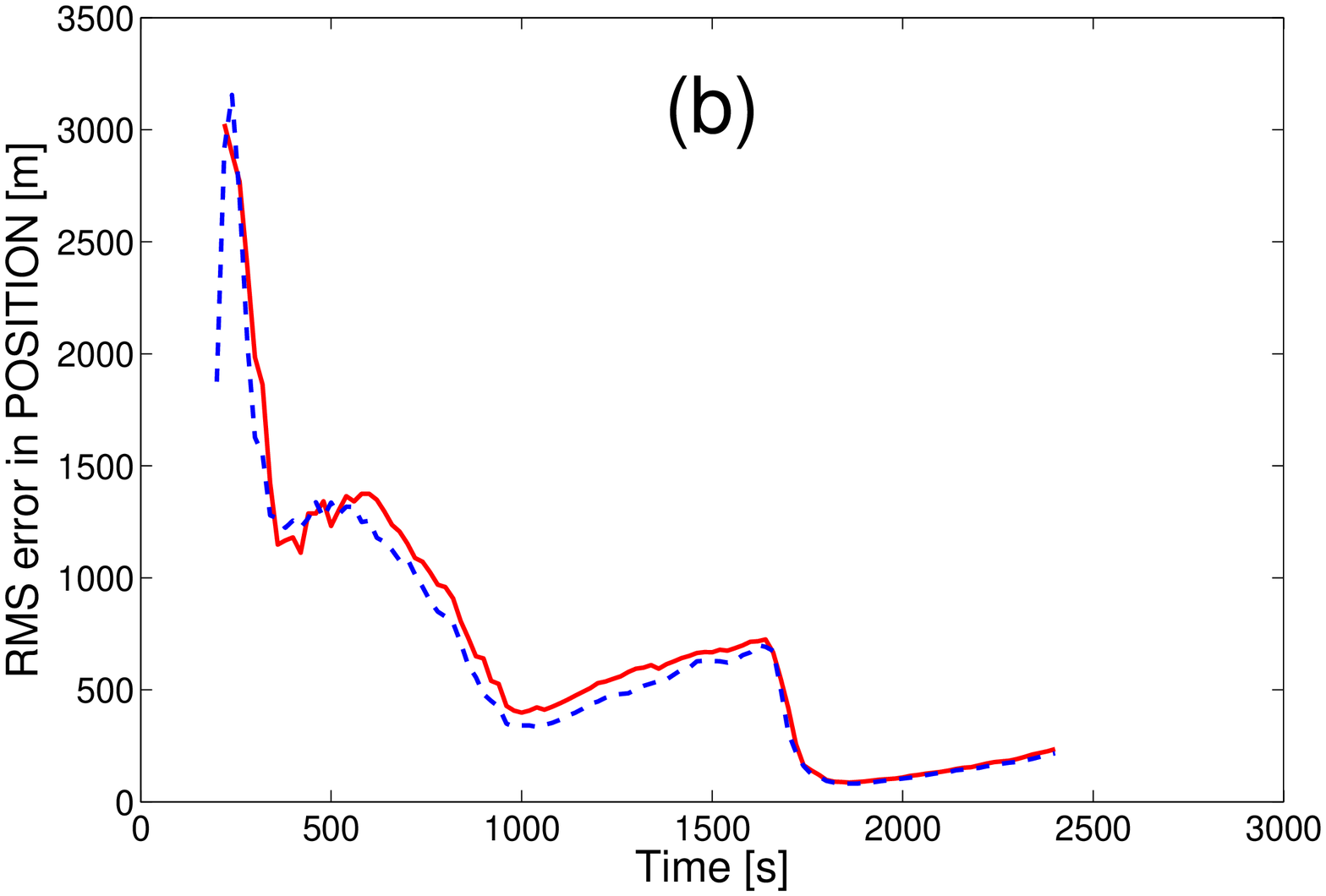}\hspace{2mm}
\includegraphics[height=4.4cm]{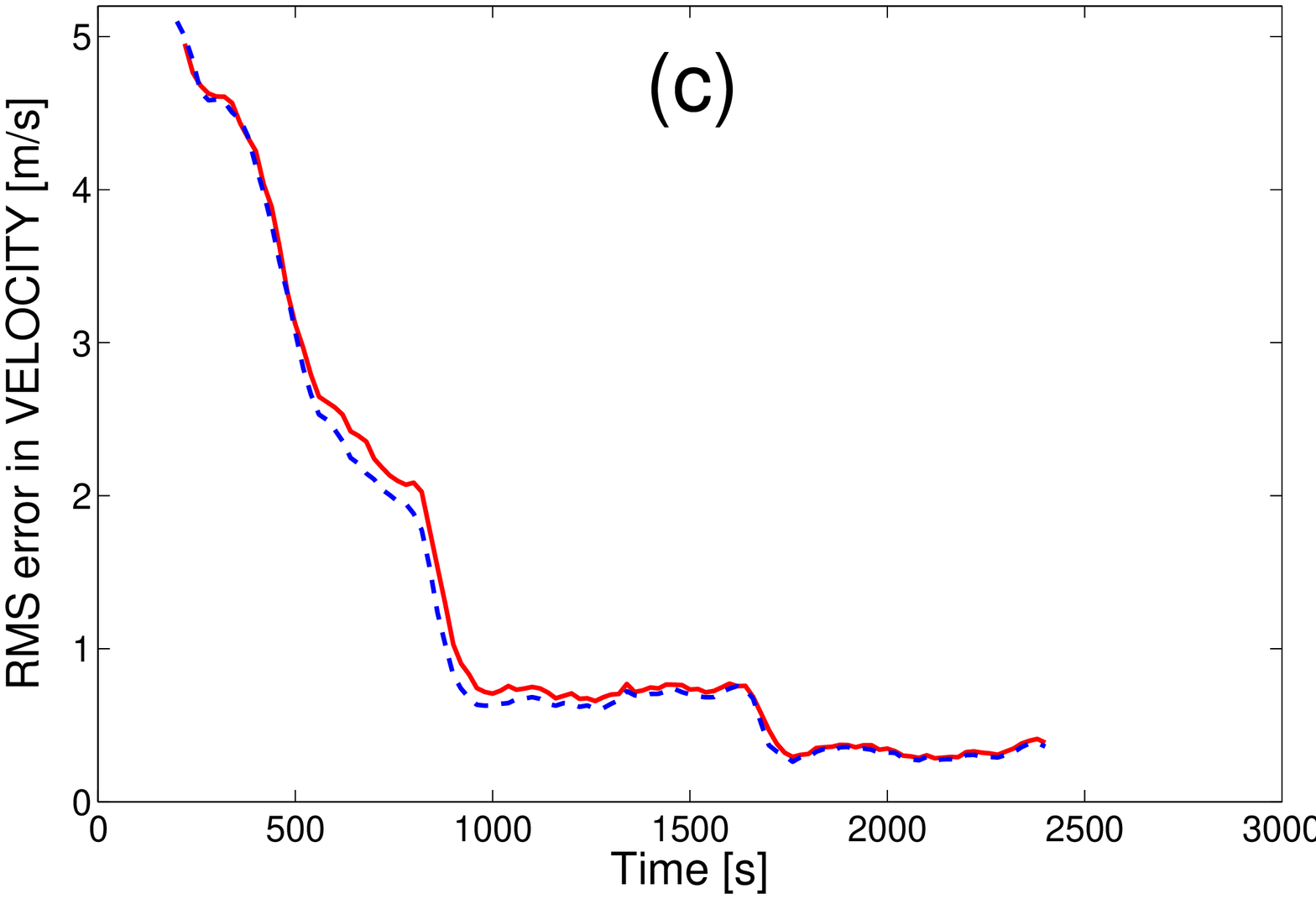}}
 \caption{Bernoulli particle filter for joint detection and tracking using bearings-only measurements
 ($p_{\sD}=0.95$; uniformly distributed clutter with $\lambda=1$ per scan; $\sigma_w=0.3^o$): (a) average
 probability of existence $r_{k|k}$ (dashed line is ideal); (b) RMS error in position; (c) RMS error in velocity (the dashed lines in (b)
 and (c) are the copies from Fig. \ref{f:1}, obtained with $p_{\sD}=1$ and
 $\lambda=0$).}
  \label{f:2}
\end{figure}

Fig. \ref{f:2} shows the results obtained by averaging over 100 Monte
Carlo runs of the Bernoulli particle filter. Fig. \ref{f:2}.(a)
displays the average estimated probability of existence $r_{k|k}$,
versus time. The true existence is indicated by the dashed blue
line. Figs. \ref{f:2}.(b) and (c) present the RMS errors in estimated
target position and velocity, respectively. The EAP estimates were
computed from $\{\xb^{(i)}_{k}\}_{1\leq i \leq N}$ only at time
steps when $r_{k|k}>0.2$. The dashed blue lines in Figs. \ref{f:2}.(b) and (c) are the copies of
the corresponding RMS errors, obtained in Sec. \ref{s:bot1} (see
Fig. \ref{f:1}) using the standard particle filter under ideal conditions (i.e. with perfect knowledge of target existence and measurements
obtained with perfect detection, that is with $p_{\sD}=1$,
$\lambda=0$).
Overall, we can observe a remarkably accurate
performance of the Bernoulli particle filter: it is only marginally worse than the performance under
ideal conditions.

\section{PHD particle filters}
\label{s:phd}

The RFS Bayes-optimal filter propagates the multi-object PDF
$f_{k|k}(\Xb_{k}|\Zb_{1:k})$, defined on the space $\mathcal{F(X)}$
of finite subsets of the single-object space $\mathcal{X}$. Since
even for a relatively small number of objects it becomes cumbersome
to work on $\mathcal{F(X)}$, several principled approximations of
the RFS Bayes-optimal filter have been proposed in the context of
various measurement models. The most popular among them is the
probability hypothesis density (PHD) filter, derived by Mahler for
the standard measurement model in \cite{mahler_phdf_03}. Instead of
propagating the posterior multi-object PDF of $\Xb_k$ over time, the PHD
filter propagates its first-order statistical moment:
the intensity function or PHD (see (\ref{eq:def-d})), which is defined on
the space $\mathcal{X}$. As a simple and fast
multi-object Bayes (sub-optimal) filter, whose computational
complexity grows (only) linearly with the number of objects, the PHD
filter has quickly become very popular among researchers. This
resulted in numerous practical applications, such as passive radar
\cite{tobias_05}, sonar \cite{clark_sonar}, computer vision
\cite{maggio_08},  traffic monitoring and road mapping
\cite{lundquist2011road,canaud2013probabilty}, robotic navigation
and mapping \cite{adams2014slam,lee2014slam}, cell microscopy
\cite{rezatofighi_13}, to name a few.

The intensity function is in general a very crude approximation of the
multi-object PDF. (Recall, the multi-object PDF can be recovered from its
intensity function only for a Poisson RFS).  Hence Mahler
subsequently introduced the Cardinalised PHD filter
\cite{mahler_cphdf_07}, which propagates jointly the intensity
function and the cardinality distribution of the multi-object PDF. The
Cardinalised PHD filter improves both the estimate of the number of
objects and the accuracy of the individual state estimates
\cite{ristic2012adaptive}, but comes at an increased complexity.
In this paper we focus only on the PHD particle filter
for the standard measurement model.

\subsection{Formulation of the PHD filter}

The posterior intensity function (the posterior PHD) at time $k$,
$D_{k|k}(\xb|\Zb_{1:k})$, is abbreviated for simplicity to
$D_{k|k}(\xb)$. Recall from Sec. \ref{s:rfs} that the expected number
of objects present at time $k$ can be estimated from $D_{k|k}(\xb)$
as $\nu_{k|k} = \int_{\mathcal{X}} D_{k|k}(\xb)\,d\xb$, where
$\nu_{k|k}\in \textsf{R}$.

Assuming that the PHD function at $k-1$ is available, the prediction
equation of the PHD filter is given by \cite{mahler_phdf_03}:
\begin{equation}
D_{k|k-1}(\xb) = \gamma_{k|k-1}(\xb) + p_{\sS} \int
\pi_{k|k-1}(\xb|\xb')\,D_{k-1|k-1}(\xb')\,d\xb' \label{e:pred}
\end{equation}
where $\gamma_{k|k-1}(\xb)$ is the PHD of the RFS of object births
between time $k-1$ and $k$. Following the measurement-driven design
of the birth density in Sec. \ref{s:bpf}, we model
$\gamma_{k|k-1}(\xb) = p_{\sS}\int
\pi_{k|k-1}(\xb|\xb')\,\gamma_{k-1}(\xb')\,d\xb'$, where assuming a
Poisson birth RFS,  $\gamma_{k-1}(\xb) = \nu_b
\,b_{k-1}(\xb|\Zb_{k-1})$. Here $\nu_b$ is the expected number of
object births between time $k-1$ and $k$ (a design parameter,
typically small, e.g. $0.1$) and $b_{k-1}(\xb|\Zb_{k-1})$ is the
birth density (\ref{e:birth_den}).

Upon receiving the measurement set $\Zb_{k}$ at time $k$, the update
step of the PHD filter is computed according to:
\begin{equation}
D_{k|k}(\xb) = \left[1-p_{\sD} + \sum_{\zb\in\Zb_{k}}
\frac{p_{\sD}g_{k}(\zb|\xb)}{\kappa_{k}(\zb) + p_{\sD} \int
g_{k}(\zb|\xb)D_{k|k-1}(\xb) d\xb}\right] \,D_{k|k-1}(\xb)
\label{e:updb}
\end{equation}
where $\kappa_{k}(\zb)$ is the PHD of the clutter RFS at time $k$.
Assuming clutter is a Poisson RFS, see (\ref{e:clutter_PDF}),  its
PHD is $\kappa_{k}(\zb) = \lambda\, c(\zb)$. Note that in
(\ref{e:updb}) for simplicity we assume that the probability of
detection $p_{\sD}$ is independent of the state. The case
$p_{\sD}(\xb)$ is a straightforward generalisation.

It can be easily verified that in the absence of clutter and target
birth, with $\Zb_k=\{\zb\}$, $p_{\sS}=p_{\sD}=1$, the PHD filter
equations (\ref{e:pred})-(\ref{e:updb}) reduce to the standard Bayes
filter equations (\ref{e:pred_eq})-(\ref{e:upd_eq}).

The PHD filter recursion (\ref{e:pred})-(\ref{e:updb}) is
initialised with $D_{0|0}(\xb)$.  In the absence of any prior, we
can set $D_{0|0}(\xb)=0$, meaning that initially there are no
objects in the surveillance volume.

\subsection{The particle method applied to PHD filtering}

In the general nonlinear/non-Gaussian context,  the PHD filter
cannot be solved analytically, and is typically implemented using
the particle method. The conceptual framework for an efficient
particle PHD filter implementation has been cast in \cite{vo_05}, where
the proposal (importance) densities for drawing particles need to
depend on the latest measurement set $\Zb_k$. How to construct
these importance densities has been a topic of intensive research
in the last decade, see \cite{Whiteley_aes},
\cite{ristic2012adaptive}, \cite{yoon_12}, \cite{high_speed_13}. The
method described below mainly follows \cite{ristic_icassp_15}.

The PHD particle filter propagates through time the particle system
$\mathcal{P}_{k}\equiv \{w_{k}^{(i)},\xb_{k}^{(i)}\}_{1\leq i \leq
N_{k}}$, which approximates $D_{k|k}(\xb)$. Note that
$w_{k}^{(i)}\geq 0$ and $\sum_{i=1}^{N_{k}} w_{k}^{(i)}=
\widehat{\nu}_{k|k} \geq 0$ is an estimate of the expected number of
objects $\nu_{k|k} = \int_{\mathcal{X}} D_{k|k}(\xb)\,d\xb$. The
pseudo-code of a PHD particle filter algorithm is given in
Alg. \ref{a:phd}. This code is general enough to be used with any
type of a particle filter (e.g. the auxiliary PF, the
 exact particle flow filter, the PF with progressive correction, see Sec. \ref{s:primer}),
naturally including the bootstrap filter. In addition, at each time
step it produces an estimate of the multi-object state
$\widehat{\Xb}_k$ and, if required, the probability of existence for
each $\xb\in\widehat{\Xb}_k$.

\begin{algorithm}[ht]
\caption{ Pseudo-code of a PHD particle filter} {\footnotesize
\begin{algorithmic}[1]
\Function{PHD Particle Filter}{} \State Input:
$\mathcal{P}_{k-1}\equiv\{w_{k-1}^{(i)},\xb_{k-1}^{(i)}\}_{1\leq i
\leq N_{k-1}}$, $\Zb_k$, $\Zb_{k-1}$
   \State Draw: $\xb^{(i)}_{k-1} \sim
b_{k-1}(\xb|\Zb_{k-1})$ for $i=N_{k-1}+1,\dots,N_{k-1}+B_{k-1}$
  \State Set weights:  $w_{k-1}^{(i)} = \nu_b/B_{k-1}$ for
  $i=N_{k-1}+1,\dots,N_{k-1}+B_{k-1}$.
\State Draw a sample: $\xb^{(i)}_{k|k-1} \sim
\pi_{k|k-1}(\xb_k|\xb^{(i)}_{k-1})$  for $i=1,\dots,N_{k-1}+B_{k-1}$
\State Predict weights: $w^{(i)}_{k|k-1} = p_{\sS} w_{k-1}^{(i)}$
for  $i=1,\dots,N_{k-1}+B_{k-1}$.
 \State
Partition $\{w_{k|k-1}^{(i)},\xb_{k|k-1}^{(i)}\}_{1\leq i \leq
N_{k-1}+B_{k-1}}$ to form clusters $\mathcal{C}_{k|k-1}(\zb)$,
$\forall \zb \in \Zb_k \cup \emptyset$ \State Initialise:
$\mathcal{P}_k=\emptyset$, $\widehat{\Xb}_k=\emptyset$ \For {every
$\zb \in \Zb_k$}
 \If
{$\mathcal{C}_{k|k-1}(\zb)\neq\emptyset$} \State
$\mathcal{C}_{k}(\zb) = PFU[\mathcal{C}_{k|k-1}(\zb),\zb]$ \Comment
See Alg. \ref{a:pfu} \State $\mathcal{P}_{k} = \mathcal{P}_{k}\cup
\mathcal{C}_{k}(\zb)$ \State  Compute existence prob. $p_{e}(\zb)$
for  $\mathcal{C}_{k}(\zb)$ \If {$p_{e}(\zb)>\eta$}, \Comment $\eta$
is reporting threshold \State Compute EAP estimate $\hat{\xb}_{k}$
from $\mathcal{C}_{k}(\zb)$ \State $\widehat{\Xb}_{k} =
\widehat{\Xb}_{k} \cup \{\hat{\xb}_{k}\}$ \EndIf \EndIf \EndFor \For
{every pair $(w_{k|k-1},\xb_{k|k-1})\in
\mathcal{C}_{k|k-1}(\emptyset)$} \If {$w_{k|k-1}
> \xi$} \Comment $\xi$ is threshold  \State $\mathcal{P}_{k} = \mathcal{P}_{k}\cup
\{((1-p_{\sD})w_{k|k-1},\xb_{k|k-1})\}$\EndIf \EndFor \State Output:
$\mathcal{P}_{k}\equiv\{w_{k}^{(i)},\xb_{k}^{(i)}\}_{1\leq i \leq
N_{k}}$; $\widehat{\Xb}_{k}$ \EndFunction
\end{algorithmic}
} \label{a:phd}
\end{algorithm}

In line 3, $B_{k-1}=N_m\cdot |\Zb_{k-1}|$ particles, for potentially
newborn targets, are drawn from the birth density
$b_{k-1}(\xb|\Zb_{k-1})$ defined in (\ref{e:birth_den}). These
particles, indexed by  $i=N_{k-1}+1,\dots,N_{k-1}+B_{k-1}$,  are
given equal weights,  $w_{k-1}^{(i)} = \nu_b/B_{k-1}$ (see line 4),
so that their sum is
$\sum_{i=N_{k-1}+1}^{N_{k-1}+B_{k-1}}w_{k-1}^{(i)} = \nu_b$.
Implementation of (\ref{e:pred}) then follows in lines 4-6: both
{\em persistent} and {\em newborn} target particles are propagated
through the transitional density in line 5, while their weights are
multiplied by $p_{\sS}$ (line 6).

Line 7 in Alg. \ref{a:phd} is very unusual and needs a detailed
explanation. Previous approaches to PHD particle filtering
\cite{vo_05}, \cite{ristic2012adaptive} treated the term in the
square brackets on the RHS of (\ref{e:updb}) as a pseudo-likelihood.
Then the weights of all predicted particles, obtained after line 6,
i.e. $\{w_{k|k-1}^{(i)},\xb_{k|k-1}^{(i)}\}_{1\leq i \leq
N_{k-1}+B_{k-1}}$ were updated directly using (\ref{e:updb}) as
follows:
\begin{equation}
\tilde{w}_{k|k}^{(i)} = \left[1-p_{\sD} + \sum_{\zb\in\Zb_{k}}
\frac{p_{\sD}g_{k}(\zb|\xb^{(i)}_{k|k-1})}{\kappa_{k}(\zb) + p_{\sD}
\sum_{i=1}^{N_{k-1}+B_{k-1}}
g_{k}(\zb|\xb_{k|k-1}^{(i)})w_{k|k-1}^{(i)}}\right]w_{k|k-1}^{(i)}
\label{e:phd_w}
\end{equation}
for $i=1,\dots,N_{k-1}+B_{k-1}$. This would be followed by
resampling (using normalised weights) and possibly particle
diversification using for example the MCMC move step. This approach,
referred to as the pseudo-likelihood update (PLU), will be
demonstrated for the sake of comparison. It suffers from two problems.
First, the particles representing the targets that were undetected at time $k$
would not survive the resampling step. As a result, these targets
would have to be re-initailsed when detected at a future time,
which is very inefficient for applications such as bearings-only
filtering (due to the un-observability of target range). Second,
estimation of the multi-target state, i.e. $\widehat{\Xb}_k$, from
the particle system after the described update step would be
cumbersome. The standard approach \cite{clark_05},
\cite[p.623]{mahler_07}, \cite[Sec.9.6]{mahler_14} is to first
estimate the number of targets as the nearest integer value
corresponding to the sum of the weights in (\ref{e:phd_w}), followed
by partitioning of the particles using a data clustering algorithm.
Finally, the mean values of the clusters of particles become the EAP
estimates included in $\widehat{\Xb}_k$. Various clustering
algorithms have been compared, such as hierarchical clustering,
k-means and the EM algorithm. Clustering in this context is
completely ad-hoc and destroys the elegance of the PHD particle
filter.

Alg. \ref{a:phd} avoids the described shortcomings. In line 7
particles are partitioned in a principled manner using the update
equation (\ref{e:updb}). Note that the RHS of (\ref{e:updb})
contains $|\Zb_k|+1$ additive terms. For each particle-measurement
pair, an additive term in the sum on the RHS of (\ref{e:updb}) can
be interpreted as the probability that measurement $\zb_j$,
$j=1,\dots,|\Zb_k|$ is due to an object in the state
$\xb^{(i)}_{k|k-1}$, $i=1,\dots,N_{k-1}+B_{k-1}$
\cite[p.78]{springer_book_13}:
\begin{equation}
P_{ij} =\frac{p_{\sD}\,g_k(\zb_j|\xb^{(i)}_{k|k-1})\,
w^{(i)}_{k|k-1}}{\kappa(\zb_j) + p_{\sD}
\sum_{\ell=1}^{N_{k-1}+B_{k-1}}g_k(\zb_j|\xb^{(\ell)}_{k|k-1})\,w^{(\ell)}_{k|k-1}}.
\label{e:ij}
\end{equation}
The remaining additive term in (\ref{e:updb}) can be interpreted as
the probability that an object in the state $\xb^{(i)}_{k|k-1}$ has
not been detected. We index this case with $j=0$, i.e. its
probability is \cite[p.78]{springer_book_13}:
\begin{equation}
P_{i0} = (1- p_{\sD})w^{(i)}_{k|k-1}. \label{e:i0}
\end{equation}
It can be easily verified that $0\leq P_{ij}\leq 1$ for all
$j=0,\dots,|\Zb_k|$.

Partitioning of
$\mathcal{P}_{k|k-1}\equiv\{w_{k|k-1}^{(i)},\xb_{k|k-1}^{(i)}\}_{1\leq
i \leq N_{k-1}+B_{k-1}}$ now proceeds based on (\ref{e:ij}) and
(\ref{e:i0}). For a particle with index $i$ in
$\mathcal{P}_{k|k-1}$, one can compute the probability distribution
over the elements of $\Zb_{k}$, plus the empty set, as follows
\begin{equation}
 p_i(j) =
\frac{P_{ij}}{\sum_{\ell=0}^{|\Zb_{k}|}P_{i\ell}}, \;\;
j=0,1,\dots,|\Zb_{k}|. \label{e:p_i}
\end{equation}
Partitioning of particles in $\mathcal{P}_{k|k-1}$ is next carried
out in a probabilistic manner as follows. For each
$i=1,\dots,N_{k-1}+B_{k-1}$ we select an index $j^i\in
\{0,1,\dots,|\Zb_k|\}$ with probability $p_i(j)$. The weighted
particle $(w_{k|k-1}^{(i)},\xb_{k|k-1}^{(i)})$ from
$\mathcal{P}_{k|k-1}$  is then assigned to cluster
$\mathcal{C}_{k+1|k}(\zb_{j^i})$. When this procedure is complete we end
up with $|\Zb_k|+1$ clusters $\mathcal{C}_{k+1|k}(\zb)$, $\forall
\zb \in \Zb_k \cup \emptyset$. Note that some of the clusters may
end up being empty  (for example, if $\zb\in\Zb_{k}$ is a false
detection, $\mathcal{C}_{k|k-1}(\zb)$ is likely to be empty).

Processing steps from line 9 to line 24 in Alg. \ref{a:phd} are
self-explanatory, except for line 11. Let us denote a component of
$D_{k|k-1}(\xb)$ which is approximated by the particles in cluster
$\mathcal{C}_{k|k-1}(\zb_j)$ by $D^{j}_{k|k-1}(\xb)$. According to
(\ref{e:updb}), the Bayes update of $D^{j}_{k|k-1}(\xb)$ using the
assigned measurement $\zb_j\in\Zb_{k}$ is as follows:
\begin{equation}
D^{j}_{k}(\xb) =
\frac{p_{\sD}\,g_k(\zb_j|\xb)\,D^{j}_{k|k-1}(\xb)}{\kappa(\zb_j) +
p_{\sD} \int g_k(\zb_j|\xb)\,D^{j}_{k|k-1}(\xb)d\xb}
\label{e:modupd}
\end{equation}
A slight modification of any standard particle filter can implement
(\ref{e:modupd}). This step is carried out in line 11 where PFU
stands for {\em particle-filter update} (to be explained later). The
output of the PFU based on $\zb\in\Zb_k$ is a cluster
$\mathcal{C}_{k}(\zb) =\{w_{k}^{(\ell)},\xb^{(\ell)}_{k}\}_{1\leq
\ell \leq L}$, which is added to the particle system
$\mathcal{P}_{k}$ in line 12.
 The sum of the updated weights in cluster $\mathcal{C}_{k}(\zb)$ is less than or equal to $1$ and represents
 the probability of existence $p_e$ (i.e. the probability  that a target, whose
 posterior PDF is approximated by $\mathcal{C}_{k}(\zb)$, exists). The probability of existence is computed in line 13, and then
 in line 14  compared to the reporting threshold $\eta$. If $p_e$ is
 above this threshold, an EAP estimate  is computed using cluster $\mathcal{C}_{k}(\zb)$, see line 15,
 to be subsequently included in the multi-object state estimate
 $\hat{\Xb}_{k}$, see line 16.

The particles of cluster $\mathcal{C}_{k|k-1}(\emptyset)$ have not
been assigned any measurement for the update,  and therefore are
treated
 differently  (see the loop between lines 20 and 24). Recall that every measurement induces the so-called
newborn target particles. If unchecked, this could potentially
result in an ever-growing number of particles over time. The
if-then clause in line 21 of Alg. \ref{a:phd} is introduced to
prevent that from happening. The particles in cluster
$\mathcal{C}_{k|k-1}(\emptyset)$ whose weights are smaller than
threshold $\xi$ are eliminated and therefore cannot propagate
further in time (e.g. those initially induced on false detections,
or those corresponding to targets that ceased to exist). The particle
elimination threshold $\xi$ (line 21) must be chosen so that the
particles on undetected but currently present (existing) objects are
not eliminated. This is particularly important in applications such
as bearings-only tracking, where target range is initially
unobservable.


The bootstrap-type PFU (line 11 in Alg. \ref{a:phd}) is described by
pseudo-code in Alg. \ref{a:pfu}. A PFU routine based on progressive
correction is discussed in \cite{ristic_icassp_15}. Let the cluster
 $\mathcal{C}_{k|k-1}(\zb)$ consist of $M$ weighted particles,
$\{w^{(m)}_{k|k-1},\xb^{(m)}_{k|k-1}\}_{1\leq m \leq M}$. According
to (\ref{e:modupd}), the weight of a predicted particle
$\xb^{(m)}_{k|k-1}$ from $\mathcal{C}_{k|k-1}(\zb)$ is updated  as:
\begin{equation}
\tilde{w}^{(m)}_{k} = \frac{p_{\sD} g_k(\zb|\xb^{(m)}_{k|k-1})
w^{(m)}_{k|k-1}}{\kappa(\zb) +  p_{\sD}
\sum_{j=1}^{M}g_k(\zb|\xb^{(j)}_{k|k-1}) w^{(j)}_{k|k-1}}
\label{e:updw}
\end{equation}
This step is carried out in line 4 of Alg. \ref{a:pfu}, followed by
resampling (using normalised weights), see the loop between lines 7
and 11. While the number of particles $M$ in the input cluster
$\mathcal{C}_{k|k-1}(\zb)$  varies from cluster to cluster,
resampling (line 7) is always carried out $L$ times, where $L$ is a
user-defined parameter. Note from line 10 (Alg. \ref{a:pfu}) that the
probability of cluster existence $p_e$ remains unchanged by the PFU
routine.

\begin{algorithm}[htbp]
\caption{Bootstrap-type PFU (line 11 of Alg. \ref{a:phd})}  {
\footnotesize
\begin{algorithmic}[1]
\Function{Bootstrap-type PFU }{} \State \textbf{Input}:
$\mathcal{C}_{k|k-1}(\zb) = \{w^{(m)}_{k|k-1}, \xb^{(m)}_{k|k-1}
\}_{m=1}^{M}$,  $\zb$ \For {$m=1,\dots,M$} \State Compute
$\tilde{w}^{(m)}_{k}$ according to (\ref{e:updw}) \EndFor \State
Compute $p_e = \sum_{m=1}^{M} \tilde{w}^{(m)}_{k}$ \For
{$\ell=1,\dots,L$} \Comment Resampling loop \State Select
$c\in\{1,\dots,M\}$ with prob. $\tilde{w}^{(c)}_{k}/p_e$ \State
$\xb^{(\ell)}_{k} \leftarrow \xb^{(c)}_{k|k-1}$ \State
$w^{(\ell)}_{k} = p_e/L$ \EndFor
 \State \textbf{Output}: Apply MCMC move and output $\mathcal{C}_{k} = \{w^{(\ell)}_{k},\xb^{(\ell))}_{k}\}_{\ell=1}^L$
\EndFunction
\end{algorithmic}}
\label{a:pfu}
\end{algorithm}

\subsection{Calibration of tracking algorithms}

Tracking algorithms are based on mathematical models, in particular
the dynamic model (for target birth/surival and motion) and the sensor
measurement model. These models typically include many parameters,
such as the target birth rate $\nu_b$, survival probability
$p_{\sS}$, the false alarm rate $\lambda$, the probability of
detection $p_{\sD}$, measurement noise variance, but also sensor
biases, various factors such as the propagation losses, receiver
gains, etc. Calibration of tracking algorithms, through estimation
of their model parameters, is an important prerequisite for their
operational deployment.

In accordance with Sec. \ref{s:calib}, let a random vector
$\thB\in\ThB\subset\mathcal{R}^{n_\theta}$ represent the static
parameter vector of interest for estimation/calibration. Again we
will indicate this in notation as
$\Pi_{k|k-1}(\Xb_k|\Xb_{k-1},\thB)$, for the FISST transitional
density and $\varphi_k(\Zb_k|\Xb_k, \thB)$, for the FISST likelihood
function. The problem is to estimate the posterior density
$p(\thB|\Zb_{1:k})\propto \varrho(\Zb_{1:k}|\thB)\,
 p(\thB)$, given the prior $p(\thB)$. Similarly to the
arguments presented in Sec. \ref{s:calib}, the complicating factor is
that the likelihood function $\varrho(\Zb_{1:k}|\thB)$ cannot be
expressed in closed-from. However, note that, similar to
(\ref{e:factor}), we can apply a decomposition:
\[ \varrho(\Zb_{1:k}|\thB) = \varrho(\Zb|\thB)\prod_{t=2}^k
\varrho(\Zb_t|\Zb_{1:t-1},\thB) \] where
\[ \varrho(\Zb_t|\Zb_{1:t-1},\thB) = \int \varphi_k(\Zb_k|\Xb_k, \thB)
f(\Xb_k|\Zb_{1:k-1},\thB)\delta \Xb_k\] The key observation here is
that the conditional likelihood $\varrho(\Zb_t|\Zb_{1:t-1},\thB)$ can
be computed as a by-product of the PHD filter, see
\cite[Eq.(116)]{mahler_phdf_03}:
 \begin{align}
\varrho(\Zb_{k}|\Zb_{1:k-1},\thB)  \propto &
  \exp \left\{-p_{\sD}(\thB)\, \int
D_{k|k-1}(\xb|\Zb_{1:k-1},\thB)d\xb\right\} \times \nonumber \\
&  \prod_{\zb\in\Zb_k} \left(\kappa(\zb|\thB)+ p_{\sD}(\thB)\int
g_k(\zb|\xb,\thB)D_{k|k-1}(\xb|\Zb_{1:k-1},\thB) d\xb \right)
\label{e:lik_meas}
\end{align}
Using the PHD particle filter, (\ref{e:lik_meas}) can be estimated
from the particle system
$\mathcal{P}_{k|k-1}\equiv\{w_{k|k-1}^{(i)},\xb_{k|k-1}^{(i)}\}_{1\leq
i \leq N_{k-1}+B_{k-1}}$, available after line 6 in Alg. \ref{a:phd},
as follows:
\begin{align}
\widehat{\ell}(\Zb_{k}|\Zb_{1:k-1},\thB)  \propto &
  \exp \left\{-p_{\sD}(\thB)\, \sum_{i=1}^{N_{k-1}+B_{k-1}} w^{(i)}_{k|k-1}\right\} \times \nonumber \\
&  \prod_{\zb\in\Zb_k} \left(\kappa(\zb|\thB)+
p_{\sD}(\thB)\sum_{i=1}^{N_{k-1}+B_{k-1}}
g_k(\zb|\xb^{(i)}_{k|k-1},\thB)w^{(i)}_{k|k-1} \right)
\label{e:lik_meas_pf}
\end{align}
Now one can apply any of the standard Monte Carlo parameter
estimation algorithms  (e.g. MCMC) to estimate the posterior
$p(\thB|\Zb_{1:K})$. Note that in every MCMC iteration, for the
proposed value of $\thB$, it is necessary to run the PHD particle
filter in order to compute (\ref{e:lik_meas_pf}). The factor of
proportionality in (\ref{e:lik_meas_pf}) cancels out and hence is
irrelevant.

The described method has been demonstrated  in the context of sensor
bias calibration, see \cite{calib_13}. However, its impact is much
broader. Suppose the calibration parameter is a stochastic dynamic
process whose evolution is independent of $\Xb_k$. Then based on the
factorization of the posterior PHD function:
$D_{k|k}(\xb_k,\thB_k|\Zb_{1:k}) =
D_{k|k}(\xb_k|\thB_k,\Zb_{1:k})\,p_{k|k}(\thB_k|\Zb_{1:k})$, one can
apply a hierarchical particle method as follows: the standard
particle filter on the $\ThB$ space, and the PHD particle filter on the
$\mathcal{X}|\ThB$ space. This approach was demonstrated in the
context of joint tracking and time-varying bias estimation in
\cite{icassp_12_sr}. Moreover, if $D_{k|k}(\xb_k|\thB_k,\Zb_{1:k})$
is analytically tractable (e.g. linear/Gaussian target tracking),
then one can effectively apply a Rao-Blackwellised formulation. This
is the essence of the algorithms proposed for: tracking groups of
targets~\cite{swainspie10}, tracking an extended target
~\cite{swainfusion10} and simultaneous localisation and mapping
(SLAM) ~\cite{mullane_11,lee2014slam,lee2013slam}.

\subsection{Demonstration: Bearings-only  multi-object filtering}
\label{s:demo_phd}

\begin{figure}[tbh]
\centerline{\includegraphics[height=5cm]{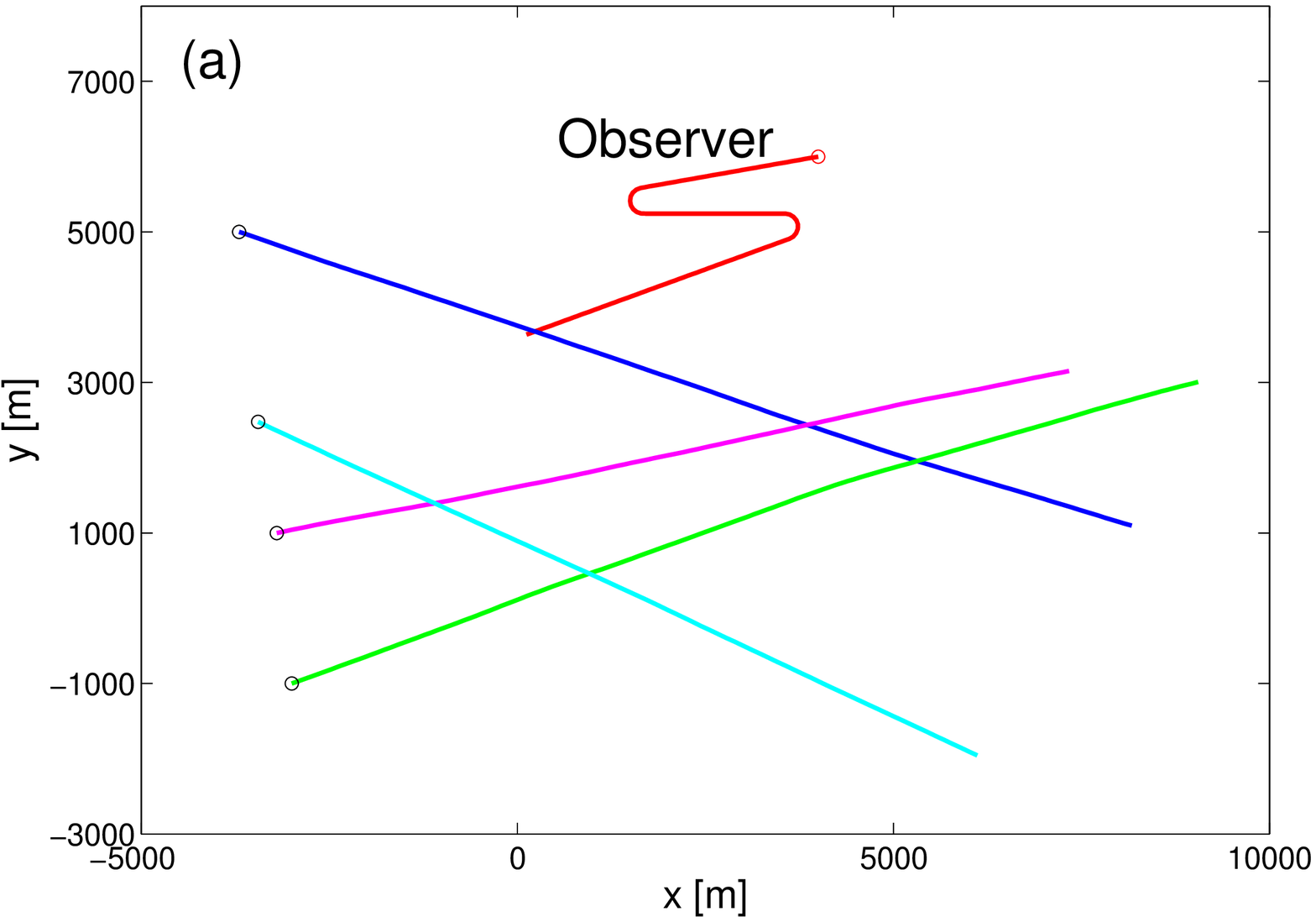}}
\centerline{\includegraphics[height=6cm]{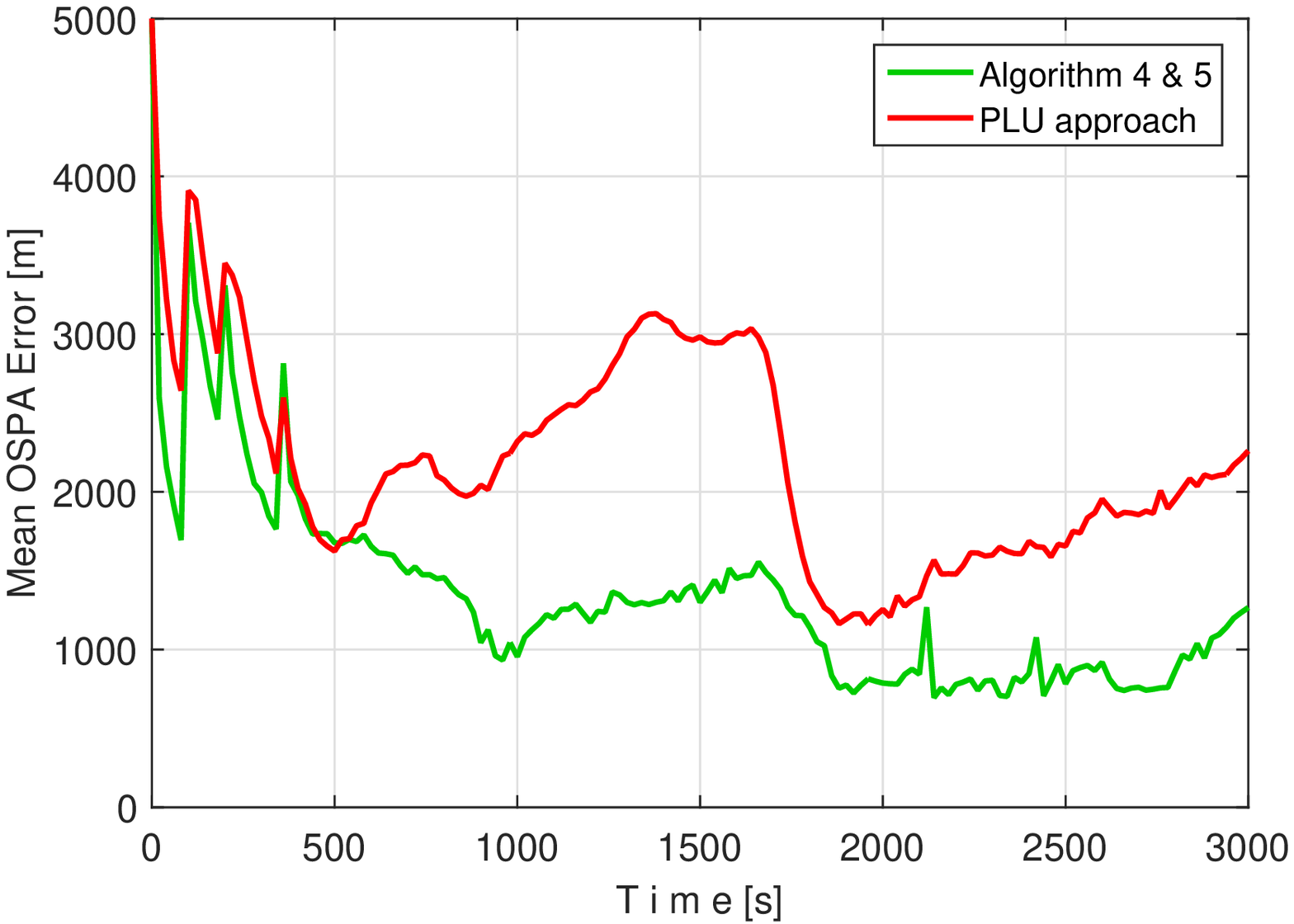}}
 \caption{Bearings-only multi-target filtering results: (a) the simulated scenario (observer and four targets, circles indicating the starting points);
 (b) The mean OSPA error (which captures both the cardinality error and the positional error) for two PHD particle filters.}
  \label{f:phd}
\end{figure}

The performance of the described PHD particle filter is demonstrated
using the simulated scenario taken from \cite{beard_15}, which
consist of four targets and lasts $3000$[s]. The geometry is shown
in Fig. \ref{f:phd}.(a). The circles on each trajectory indicate the
starting points. One target is present at the beginning, with
another three arriving during the first $400$[s]. Three targets are
terminating in the final $900$[s]. The bearings of all four targets
cross one another in the middle of the observation interval. We
assume an ideal sensor with infinite resolution.

In order to measure the statistical error of multi-object filtering
at each time step we need a distance metric between two finite sets
of objects: the ground truth $\Xb_k$ and its estimate
$\widehat{\Xb}_k$. The optimal subpattern assignment (OSPA) distance
\cite{SchuhmacherOSPA07} has become a widespread metric for this
purpose as it captures both the difference in the cardinality of the
two finite sets and the positional  error of their respective
elements. The OSPA distance has two parameters: (1) the cut-off value, which
we adopt as $c=5$[km]; (2) the order parameter of the Euclidian base
distance (in position), which we adopt as $p=2$.

Fig. \ref{f:phd}.(b) shows the OSPA  error averaged over 500 Monte
Carlo runs, for two PHD particle filters: the one described by Algs.
\ref{a:phd} and \ref{a:pfu}, and the alternative, based on the
pseudo-likelihood update step \cite{ristic2012adaptive}. The
parameters used in simulations were: $p_{\sD}=0.95$; Poisson clutter
with $\lambda=1$ and
 $c(z)=(2\pi)^{-1}$[rad$^{-1}$]; $N_m=2500$,
$p_b=0.01$, $p_{\sS}=0.98$; $\sigma_w = 1^o$, $T=20$[s],
$\sigma_v=0.005$ [m/s$^2$], $r_{\max}=10000$[m], and $N=5000$
particles. Clearly, the PHD particle filter described by Algs.
\ref{a:phd} and \ref{a:pfu}, which includes the partitioning of
particles,  performs much better than the alternative based on PLU.

\section{Labelled RFS Bayes tracking filters}
\label{s:lrfs}

Since the introduction of RFS Bayes filters (optimal and
suboptimal), there have been numerous attempts to apply them to
multi-target tracking problems, e.g. \cite{clark_05}, \cite{lin_06},
\cite{maggio_08},  \cite{yang_12}. These heuristic methods typically
label the targets and apply some form of data association to keep
track of the labels as time progresses. Mahler's view
\cite{mahler_07} was that the most natural and therefore convenient RFS for multi-target
tracking is the multi-Bernoulli RFS, introduced in
Sec. \ref{s:common}. His initial formulation led to the development
of an approximation of the RFS Bayes-optimal filter referred to as the
CBMeMBer filter \cite{cbmember_09}. Assuming the multi-object PDF
$f_{k|k}(\Xb_k|\Zb_{1:k})$  is multi-Bernoulli, the CBMeMBer filter
propagates the pairs $\big(r_{k|k}^j,p_{k|k}^j(\xb)\big)$,
$j=1,\dots,M_k$ over time via the prediction and update steps. While
 the prediction step (\ref{e:pred_PDF}) for the CBMeMBer filter can be derived in
closed-form exactly, the update step (\ref{e:upd_PDF}) 
formulation is based on a (fairly un-intuitive) approximation.

The breakthrough was made in \cite{vo2013labeled} with the
formulation of {\em labelled} random finite sets.
While preserving the mathematical rigor of the RFS framework, the labelled RFS
concept allows the assignment of a distinct label to each object (target)
that appears in the surveillance volume during the observation time.
The history of the state evolution for each object (i.e. its
trajectory) can then be naturally identified. More importantly, this
framework allows the formulation of an exact closed-from Bayes-optimal
multi-target tracking filter \cite{vo2014labeled}. Related relevant recent papers include \cite{williams_15, correa_15}.

\subsection{Labelled RFS}

Suppose the elements of an RFS $\Xb$ on $\mathcal{X}$ are uniquely
tagged by an (unobserved) label drawn from a discrete countable
space of labels  $\mathcal{L}$. A labelled RFS, denoted
$\mathbb{X}$, is an RFS on the product space
$\mathcal{X}\times\mathcal{L}$, such that each realisation of
$\mathbb{X}$ has distinct labels. We express distinctness of labels
mathematically by introducing the {\em distinct label indicator}
function $\Delta(\mathbb{X})$. Specifically, if we assume that
$\Xm=\{(\xb_1,\ell_1),\cdots,(\xb_n,\ell_n)\}$, then the  distinct label indicator is defined as
\[\Delta(\mathbb{X})=\delta[\,|\mathbb{X}|,|\mathfrak{L}_{\mathbb{X}}|],\]
where $\mathfrak{L}_{\mathbb{X}}=\{\ell_1,\cdots,\ell_n\}$ is the
set of unique labels in $\mathbb{X}$ and  $\delta[a,b] $ is the
Kroneker delta function, defined earlier. 
By definition,  a realisation of a labelled RFS $\mathbb{X}$ always
satisfies $\Delta(\mathbb{X})=1$. Labelled RFSs include,
for example, the labelled Poisson RFS, and the labelled
multi-Bernoulli RFS, see \cite{vo2013labeled}. In order to distinguish the
distributions and statistics of labelled versus unlabelled random variables (both vectors and finite sets),  functions of labelled random variables will be denoted with a tilde sign. For example, a PDF  on the space $\mathcal{X}\times\mathcal{L}$ is denoted $\tilde{p}$, while a PDF on $\mathcal{F}(\mathcal{X}\times\mathcal{L})$ is denoted $\tilde{f}$.

For an analytical formulation of the RFS Bayes optimal tracking
filter, of particular interest is a general class of
labelled RFS,  referred to as the {\em generalised labelled
multi-Bernoulli} (GLMB) RFS \cite{vo2013labeled}. The PDF of a GLMB RFS $\mathbb{X}$, with single object state space  $\mathcal{X}$ and discrete-label space $\mathcal{L}$
is given by:
\begin{equation}
\tilde{f}(\mathbb{X}) = \Delta(\mathbb{X})\;\sum_{j\in J}
\phi^{j}(\mathfrak{L}_{\mathbb{X}})\prod_{(\xb,\ell)\in\Xm}\,\tilde{p}^j(\xb,\ell)
\label{e:GLMB_Density}
\end{equation}
where $J$ is an arbitrary index set, and $\phi^j$ and $\tilde{p}^j$
satisfy
\[ \sum_{L\subseteq \mathcal{L}}\sum_{j\in J} \phi^j(L) = 1\mbox { and
} \int_{\xb\in\mathcal{X}}\tilde{p}^j(\xb,\ell)d\xb=1, \;\forall
\ell\in\mathcal{L}. \]

This RFS has a
remarkable property that its PDF is a conjugate prior for the
standard multi-object likelihood function (\ref{e:mo_lik}) and is closed under
the Chapman-Kolmogorov prediction with the standard multi-object transitional
model.

The GLMB is a very general and flexible RFS model, but its application
to multi-target tracking problems is more easily demonstrated by considering
a more specific type of GLMB RFS, which was also proposed in \cite{vo2013labeled}.
A $\delta$-generalised labelled multi-Bernoulli ($\delta$-GLMB)
RFS with state space $\mathcal{X}$ and label space $\mathcal{L}$,
is a GLMB RFS with density of the same form as (\ref{e:GLMB_Density}),
with the following substitutions
\begin{align}
J & =\mathcal{F}\left(\mathcal{L}\right)\times\Xi,\\
\phi^{j}\left(L\right) & =\phi^{I,\xi}\left(L\right)=\phi^{I,\xi}\delta\left[I,L\right],\\
\tilde{p}^{j} & =\tilde{p}^{I,\xi}=\tilde{p}^{\xi},
\end{align}
where $\delta[X,Y]$ is a generalisation of the Kroneker delta for sets (i.e. $\delta[X,Y]=1$ if $X=Y$ and zero otherwise), and $\Xi$ is a discrete space. Thus, a $\delta$-GLMB RFS is distributed
according to
\begin{align}
\tilde{f}\left(\mathbb{X}\right) & =\Delta\left(\mathbb{X}\right)\sum_{\left(I,\xi\right)\in\mathcal{F}
\left(\mathcal{L}\right)\times\Xi}\phi^{I,\xi}\delta\left[I,\mathfrak{L}_{\mathbb{X}}\right]
\prod_{\left(\mathbf{x},\ell\right)\in\mathbb{X}}\tilde{p}^{\xi}\left(\mathbf{x},\ell\right).
\label{e:Delta_GLMB}
\end{align}

In a $\delta$-GLMB, the sum is taken over the Cartesian product between
the space of finite subsets of $\mathcal{L}$, and the discrete space
$\Xi$. In tracking applications, each subset of $\mathcal{L}$ represents
a set of target labels, while $\Xi$ represents
the space of measurement-to-label association histories. Hence, a
particular element $\left(I,\xi\right)\in\mathcal{F}\left(\mathcal{L}\right)\times\Xi$
can be considered to be the hypothesis that the set of currently existing
targets are those with labels $I$ and association history $\xi$.
The weight $\phi^{I,\xi}$ is the probability of this hypothesis,
and $\tilde{p}^{\xi}\left(\xb,\ell\right)$ is the PDF of the target
with label $\ell$ under the association history $\xi$.

Like the GLMB, the $\delta$-GLMB has also been shown to be closed
under the standard multi-object prediction, and a conjugate prior
with respect to the standard multi-object measurement likelihood \cite{vo2013labeled}.
This makes the $\delta$-GLMB a highly useful tool for deriving analytical
Bayes recursions for multi-object tracking problems.

\subsection{Labelled multi-object transition and likelihood models}
\label{s:Labelled_Transition}

The labelled version of the likelihood model $\varphi_k(\Zb|\mathbb{X})$ essentially has the same form as (\ref{e:mo_lik}), with only a few notes of caution. First, the probability of detection is assumed to depend on the state of the target and its label, and therefore $p_\sD$ in (\ref{e:mo_lik}) should be replaced with $\tilde{p}_{\sD}(\xb,\ell)$. Second, a target-to-measurement association is defined as a mapping  $\theta:\mathfrak{L}_{\mathbb{X}}\rightarrow\{0,1,\dots,|\Zb|\}$. Finally, in the context of labelled RFSs, we need to substitute $\xb_i$ and  $\Xb$ in (\ref{e:mo_lik}) with $(\xb_i,\ell_i)$  and $\mathbb{X}$, respectively.

The labelled version of the standard multi-target
dynamic model is more involved. Let $\mathbb{X}$ be the labelled RFS of objects at
the current time with label space $\mathcal{L}$. A particular object
$\left(\mathbf{x},\ell\right)\in\mathbb{X}$ has probability $\tilde{p}_{\sS}\left(\mathbf{x},\ell\right)$
of surviving to the next time with state $\left(\mathbf{x}_{+},\ell_{+}\right)$
and probability density $\pi\left(\mathbf{x}_{+}|\mathbf{x},\ell\right)\delta\left[\ell,\ell_{+}\right]$
(where $\pi\left(\mathbf{x}_{+}|\mathbf{x},\ell\right)$ is the single
target transition kernel), and probability, $1-\tilde{p}_{\sS}\left(\mathbf{x},\ell\right)$
of being terminated. Thus, the set $\mathbb{S}$ of surviving objects
at the next time is distributed according to
\begin{equation}
\tilde{\Pi}_{S}\left(\mathbb{S}|\mathbb{X}\right)=\Delta\left(\mathbb{S}\right)
\Delta\left(\mathbb{X}\right)
1_{\mathfrak{L}_{\mathbb{X}}}\left(\mathfrak{L}_{\mathbb{S}}\right)\prod_{\left(\mathbf{x},\ell\right)\in\mathbb{X}}\tilde{\Phi}\left(\mathbb{S};\mathbf{x},\ell\right),\label{e:Survival_Transition}
\end{equation}
where $1_A(B)$ is a generalisation of the indicator function for sets, such that $1_A(B)=1$ if $B\subseteq A$ and zero otherwise, and
\begin{equation}
\tilde{\Phi}\left(\mathbb{S};\mathbf{x},\ell\right)  =\begin{cases}
\tilde{p}_{\sS}\left(\xb,\ell\right)\,\pi(\xb_{+}|\xb,\ell), & \mbox{ if } \ell\in\mathfrak{L}_{\mathbb{S}} \\
1 - \tilde{p}_{\sS}\left(\mathbf{x},\ell\right), & \mbox{ if } \ell\not\in\mathfrak{L}_{\mathbb{S}}
\end{cases}
\end{equation}

Let $\mathbb{B}$ be the labelled RFS of newborn objects with label
space $\mathcal{B}$, where $\mathcal{L}\cap\mathcal{B}=\emptyset$.
Since the births have distinct labels, and assuming that their states
are independent, we model $\mathbb{B}$ as a labelled multi-Bernoulli
(LMB) RFS, which is distributed according to
\begin{align}
\tilde{f}_{b}\left(\mathbb{B}\right) & =\Delta\left(\mathbb{B}\right)\phi_{b}\left(\mathfrak{L}_{\mathbb{B}}\right)
\prod_{\left(\mathbf{x},\ell\right)\in\mathbb{B}}\tilde{b}\left(\mathbf{x},\ell\right),
\label{e:LMB_Birth}
\end{align}
where $\tilde{b}\left(\mathbf{x},\ell\right)$ is the single object
birth density corresponding to label $\ell$, and $\phi_{b}\left(L\right)$
is the birth weight defined by
\begin{align}
\phi_{b}\left(L\right) & =\prod_{\ell\in\mathcal{B}\setminus L}\left(1-r_{b}\left(\ell\right)\right)\prod_{\ell\in L} \;1_{\mathcal{B}}\left(\ell\right)\;r_{b}\left(\ell\right),
\label{e:LMB_Birth_Weight}
\end{align}
in which $r_{b}\left(\ell\right)$ is the existence probability of the newborn object
with label $\ell$.


The overall prediction of the multi-object state at the next time
step is the union of survivals and new births, i.e. $\mathbb{X}_{+}|\mathbb{X}=\mathbb{S}\cup\mathbb{B}$.
The label spaces $\mathcal{L}$ and $\mathcal{B}$ are disjoint, and
the states of newborn objects are independent of surviving objects,
hence $\mathbb{S}$ and $\mathbb{B}$ are independent. It can be shown \cite{vo2013labeled} that the multi-object transition can be expressed as a product of the transition density for surviving objects and the density of newborn objects:
\begin{equation}
\tilde{\Pi}\left(\mathbb{X}_{+}|\mathbb{X}\right)=
\tilde{\Pi}_{S}\left(\mathbb{X}_{+}\cap\left(\mathcal{X}\times\mathcal{L}\right)|\mathbb{X}\right)
\cdot \tilde{f}_{b}\left(\mathbb{X}_{+} \setminus \left(\mathcal{X}\times\mathcal{L}\right)\right).
\label{e:Labelled_Multi-object_Transition}
\end{equation}
Both the GLMB and $\delta$-GLMB families are closed under
the Chapman-Kolmogorov prediction with this transition \cite{vo2013labeled}.

\subsection{$\delta$-GLMB particle filter}

Suppose that the posterior multi-object density at time
$k-1$ is a $\delta$-GLMB with the label space $\mathcal{L}$, given by (\ref{e:Delta_GLMB}), i.e.
 \begin{equation}
 \tilde{f}_{k-1}(\Xm) = \Delta(\mathbb{X})\;\sum_{\left(I,\xi\right)\in\mathcal{F}
\left(\mathcal{L}\right)\times\Xi}
\phi_{k-1}^{I,\xi}\delta[I,\mathfrak{L}_{\Xm}]
\prod_{(\xb,\ell)\in\Xm}\,\tilde{p}^\xi_{k-1}(\xb,\ell)
\label{e:post_lmo_dens}
\end{equation}
 in which the distribution $\tilde{p}^{\xi}_{k-1}(\xb,\ell)$
of the target with label $\ell$ and measurement association history
$\xi$ is represented
by a weighted set of particles \[\{w_{k-1}^{(i),\xi}(\ell),\xb_{k-1}^{(i),\xi}(\ell)\}_{1\leq i\leq N^\xi_{\ell}}.\] The particle weights are normalised, i.e. $\sum_{i=1}^{N^\xi_{\ell}} w_{k-1}^{(i),\xi}(\ell) = 1$ and we can write:
\begin{align}
\tilde{p}^{\xi}(\xb,\ell) & \approx\sum_{i=1}^{N^{\xi}_{\ell}}w_{k-1}^{(i),\xi}(\ell)\;\delta\left(\xb-\xb_{k-1}^{(i),\xi}(\ell)
\right).
\end{align}

The goal of the $\delta$-GLMB particle filter is to recursively compute the
posterior $\delta$-GLMB density, conditioned on new sets of measurements as
they are received. As usual this is done by first applying the Chapman-Kolmogorov
equation with the multi-object transition model, which yields
the predicted $\delta$-GLMB density at the next observation time $k$. Bayes rule
is then applied to this prediction using the multi-object
likelihood model and the received measurement set. The result is the posterior
$\delta$-GLMB density at time $k$. The prediction and update steps are described next.

\subsubsection*{Prediction}

Let us assume that within the LMB birth model (\ref{e:LMB_Birth}),
the single-object density $\tilde{b}(\xb,\ell)$ corresponding to the target with label $\ell$
is approximated by particles $\{w_{b}^{(i)}(\ell),\xb_{b}^{(i)}(\ell)\} _{1\leq i \leq N^b_{\ell}}$.
Under the labelled multi-object transition model defined in (\ref{e:Labelled_Multi-object_Transition}),
the predicted multi-object density at time $k$ is a $\delta$-GLMB
 with the label space $\mathcal{L}_{+}=\mathcal{L}\cup \mathcal{B}$, given by \cite{vo2013labeled}
\begin{equation}
\tilde{f}_{k|k-1}(\Xm)=
\Delta(\Xm)\sum_{(I_+,\xi)\in\mathcal{F}(\mathcal{L}_+)\times \Xi}
\phi_{+}^{I_+,\xi}\delta[I_+,\mathfrak{L}_{\Xm}]
\prod_{(\xb,\ell)\in\Xm}\tilde{p}_{+}^{\xi}(\xb,\ell),
\label{e:pred_glmb}
\end{equation}
where, assuming the bootstrap-type particle filter with the importance density equal the (single-target) transitional density,
\begin{align}
\phi_{+}^{I_+,\xi}(L) & =\phi_{b}(I_+\cap\mathcal{B})\cdot \phi_{\sS}^{\xi}(I_+\cap\mathcal{L}),\label{e:GLMB_pred_weight}\\
\tilde{p}_{+}^{\xi}(\xb,\ell) & \approx
\begin{cases}
\sum_{i=1}^{N^{\xi}_{\ell}} w_{\sS}^{(i),\xi}(\ell)\,
\delta\left(\xb-\xb_{\sS}^{(i),\xi}(\ell)\right), & \mbox{ if } \ell \in \mathcal{L}\\
\sum_{i=1}^{N^b_{\ell}} w_{b}^{(i)}(\ell)\,\delta\left(\xb-\xb_{b}^{(i)}(\ell)\right),
& \mbox{ if }
\ell \in \mathcal{B}
\end{cases}
\label{e:GLMB_p_predict}\\
\xb_{\sS}^{(i),\xi}(\ell) & \sim \pi_{k|k-1}\left(\xb|\xb_{k-1}^{(i),\xi}(\ell)\right),\quad i=1,\dots,N^{\xi}_{\ell}\label{e:GLMB_proposal}
\\
w_{\sS}^{(i),\xi}(\ell) & \propto
w_{k-1}^{(i),\xi}(\ell)\;\tilde{p}_{\sS}\left(\xb_{k-1}^{(i),\xi}(\ell),\ell\right),\quad i=1,\dots,N^{\xi}_{\ell}\\
\eta_{\sS}^{\xi}(\ell) & =\sum_{i=1}^{N^{\xi}_{\ell}}w_{k-1}^{(i),\xi}(\ell)\;\tilde{p}_{\sS}
\left(\xb_{k-1}^{(i),\xi}(\ell),\ell\right),\label{e:GLMB_eta_S}\\
\phi_{\sS}^{\xi}\left(M \right) & =\left(\prod_{\ell\in M}\eta_{\sS}^{\xi}(\ell)\right)\sum_{I\subseteq\mathcal{L}}1_{I}\left(M\right)\phi_{k-1}^{\xi}
\left(I\right)
\prod_{\ell\in I \setminus M}\left[1-\eta_{\sS}^{\xi}(\ell)\right].\label{e:GLMB_surv_weight}
\end{align}

In principle, the $\delta$-GLMB prediction involves generating a new predicted
component (hypothesis) for every combination of target birth, death  and survival,
for each component in the current posterior density. In practice however,
doing so is usually infeasible because the number of combinations
becomes prohibitively large. To improve the efficiency, approximations
can be made that avoid computing unlikely components. One approach
is to use a ``shortest paths'' algorithm to generate the most likely
combinations of targets.

\begin{figure}[h]
\centerline{\includegraphics[width=0.4\columnwidth]{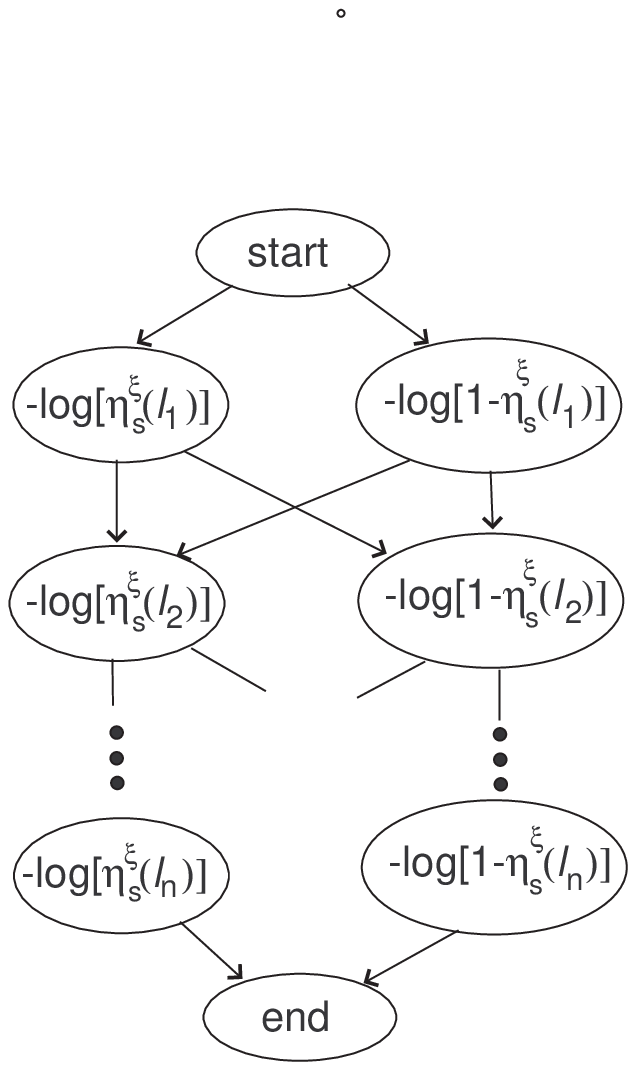}
\includegraphics[width=0.4\columnwidth]{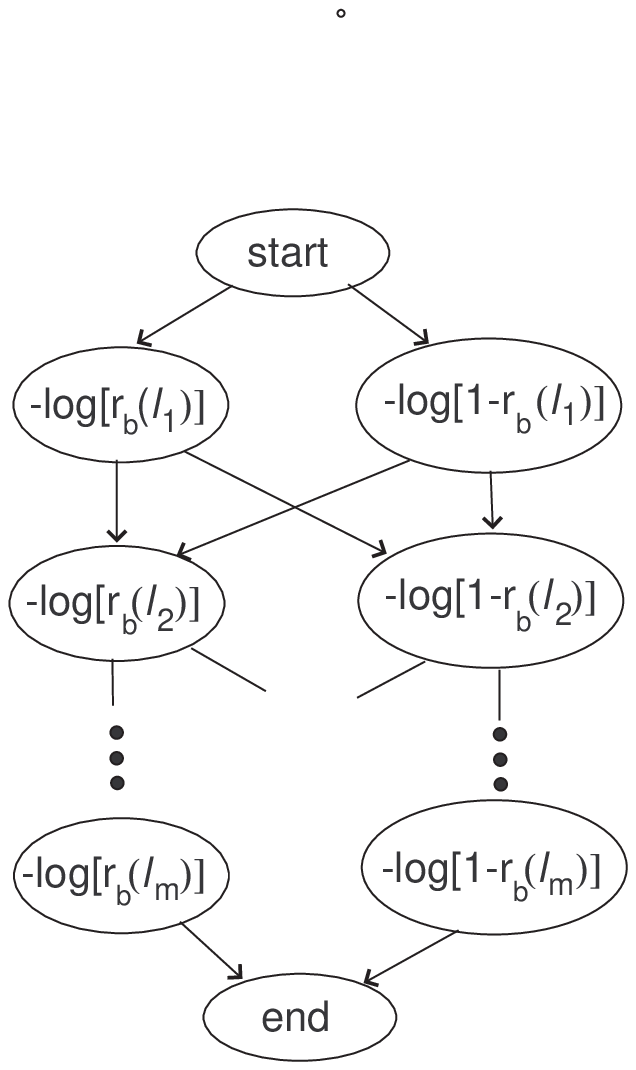}
}
\centerline{\footnotesize (a)\hspace{5.5cm}(b)}
\caption{Graph structures for $\delta$-GLMB prediction: (a) survival prediction graph; (b) birth prediction graph. Note that the labels in (a) and (b) are disjoint sets.}
\label{f:GLMB_Prediction_Graphs}
\end{figure}

This can be carried out separately for births
and survivals, using the graph structures shown in Fig. \ref{f:GLMB_Prediction_Graphs}. Each
row in the graphs in Fig. \ref{f:GLMB_Prediction_Graphs}
corresponds to a target label: the left-hand column represents target
survival/birth, and the right-hand column represents target death/non-birth.
The goal is to maximise the predicted component weights, which are
expressed in (\ref{e:GLMB_pred_weight}), (\ref{e:GLMB_surv_weight})
and (\ref{e:LMB_Birth_Weight}) as products of single-target weights.
Since the shortest paths algorithm is based on minimising a set of
additive path costs, negative logarithms are used to change the problem
from maximum-product to minimum-sum form. Paths are generated from
top-to-bottom, and only those targets corresponding to rows in which
the left-hand column was visited are included in the prediction.

Pseudo-code of the prediction step of the $\delta$-GLMB particle filter is given
in Alg. \ref{a:GLMB_Prediction}. The input to this function
in line 2 represents $\tilde{f}_{k-1}(\Xm)$ of (\ref{e:post_lmo_dens}), with $L_{k-1}^h$ being the set of target labels according to hypothesis $h=1,\dots,H$. The input also includes the number of predicted components to be generated for each component in the current density, that is $K^{h}$ for $h=1,\dots,H$. The input
in line 3 represents $\tilde{f}_{b}(\mathbb{B})$ of (\ref{e:LMB_Birth}). Note that the $\mathtt{ShortestPaths}$ function returns
a collection of pairs $\left\{ \left(L^{j},c^{j}\right)\right\} _{1\leq j\leq K}$,
where each $L^{j}$ is a list of surviving target labels, and $c^{j}$
is the corresponding path cost (the negative logarithm of
the weight).

\begin{algorithm}[htb]
\caption{The prediction step of the $\delta$-GLMB particle filter }
{ \footnotesize
\begin{algorithmic}[1]
\Function{$\delta$-GLMB particle filter: Prediction step}{}
\State Input: $\left\{L_{k-1}^{h},\phi_{k-1}^{h},
               \left\{\left\{w^{(i),h}_{k-1}(\ell),\xb^{(i),h}_{k-1}(\ell)\right\}_{1\leq i\leq N^{h}_{\ell}}
               \right\}_{\ell\in L_{k-1}^h},
               K^{h}\right\}_{1\leq h \leq H_{k-1}}$
\State Input: $\left\{ r_b(\ell),
                 \left\{ w_{b}^{(i)}(\ell),\xb_{b}^{(i)}(\ell) \right\}_{1\leq i\leq N^b_{\ell}} \right\}_{\ell\in\mathcal{B}}$,
              $K_b$
\State Construct graph $\mathcal{G}_b$ according to Figure \ref{f:GLMB_Prediction_Graphs}.(b)
\State $\left\{\left(L_b^j,c_b^j\right)\right\}_{1\leq j \leq K_b}
        = \mathtt{ShortestPaths}\left(\mathcal{B},\mathcal{G}_b,K_b\right)$
\State $m = 1$
\For{$h=1,\dots,H_{k-1}$}
  \State Construct graph $\mathcal{G}_{\sS}^h$ according to Figure \ref{f:GLMB_Prediction_Graphs}.(a)
  \State $\left\{\left(L_{\sS}^{h,j},c_{\sS}^{h,j}\right)\right\}_{1\leq j \leq K^h}
          = \mathtt{ShortestPaths}\left(L_{k-1}^h,\mathcal{G}_{\sS}^h,K^h\right)$
  \For{$j=1,\dots,K^h$}
    \For{$\beta=1,\dots,K_b$}
      \State $L_+^m = L_{b}^\beta \cup L_{\sS}^{h,j}$
      \State $\phi_+^m = \phi_{k-1}^h \exp\left(-c_b^\beta\right) \exp\left(-c_{\sS}^{h,j}\right)$
      \State Compute $\left\{
                        w_{+}^{(i),m}(\ell),\xb_{+}^{(i),m}(\ell)
                      \right\}_{1\leq i\leq N_+^{m}(\ell)}, \forall \ell \in L_+^m$
            \Comment Eqn (\ref{e:GLMB_p_predict})
      \State $m = m + 1$
    \EndFor
  \EndFor
\EndFor
\State $H_+ = m - 1$
\State Normalise weights $\left\{\phi_+^j\right\}_{1\leq j \leq H_+}$
\State Output:
       $\left\{ L_+^h, \phi_+^h,\left\{\left\{
         w_{+}^{(i),h}(\ell), \xb_{+}^{(i),h}(\ell) \right\}_{1\leq i\leq N^h_+(\ell)}\right\}_{\ell\in L^h_+} \right\}_{1\leq h \leq H_+}$
\EndFunction
\end{algorithmic}
}
\label{a:GLMB_Prediction}
\end{algorithm}

\subsubsection*{Update}

The prior multi-object density at time $k-1$ is a $\delta$-GLMB
given by (\ref{e:pred_glmb}) and represented by line 21 of Alg. \ref{a:GLMB_Prediction}. Next we wish to compute the posterior multi-object density $\tilde{f}_{k|k}(\Xm)$, conditioned on a set of received measurements $\Zb_k$.

Let $\Theta$ be the set of all mappings of target
labels to measurement indices, $\theta:\mathbb{L}\rightarrow\left\{ 0,1,...,\left|Z\right|\right\} $,
such that $\left[\theta\left(i\right)=\theta\left(j\right)>0\right]\Rightarrow\left[i=j\right]$.
Note that a given $\theta$ can map multiple labels to $0$, meaning
that there can be multiple misdetected targets. However, only one
label can be mapped to each non-zero measurement index, which means
that each measurement can only be assigned to one target. Under the
standard multi-object observation likelihood $\varphi_k(\Zb|\mathbb{X})$ (see the first paragraph of Sec. \ref{s:Labelled_Transition}), the posterior multi-object
density is a $\delta$-GLMB given by \cite{vo2013labeled}
\begin{align}
\tilde{f}_{k|k}(\Xm) & =\Delta(\Xm)
\sum_{\left(I_+,\xi\right)\in\mathcal{F}\left(\mathcal{L}_+\right)\times\Xi}\;\sum_{\theta\in\Theta
\left(\mathfrak{L}_{\mathbb{X}}\right)}\phi_{Z}^{I_+,\xi,\theta}\delta[I_+,\mathfrak{L}_{\mathbb{X}}]
\prod_{(\xb,\ell)\in\Xm}\tilde{p}^{\xi,\theta}\left(\xb,\ell|\Zb_k\right),
\end{align}
where
\begin{align}
\phi_{Z}^{I_+,\xi,\theta}(L) & = \frac{\phi_+^{I_+,\xi}(L)\prod\limits _{\ell\in I_+}\eta_{Z}^{\xi,\theta}(\ell)}{\sum\limits _{\left(J,\alpha\right)\in\mathcal{F}\left(\mathcal{L}_+\right)\times\Xi}\;\sum\limits _{\theta\in\Theta\left(J\right)}\omega^{J,\alpha}\left(J\right)\prod\limits _{\ell\in J}\eta_{Z}^{\alpha,\theta}\left(\ell\right)},\\
\tilde{p}^{\xi,\theta}(\xb,\ell|\Zb) & \approx\sum_{i=1}^{N^{\xi}_{\ell}}\frac{w_{+}^{(i),\xi}(\ell)\tilde{\psi}_{Z}
\left(\xb_{+}^{(i),\xi}(\ell),\ell;\theta\right)}{\eta_{Z}^{\xi,\theta}\left(\ell\right)}\,
\delta\left(\xb-\xb_{+}^{(i),\xi}(\ell)\right),\label{e:p_update}\\
\eta_{Z}^{\xi,\theta}(\ell) & =\sum_{i=1}^{N^{\xi}_{\ell}}w_{+}^{(i),\xi}(\ell)\tilde{\psi}_{Z}\left(\mathbf{x}_{+}^{(i),\xi}
(\ell),\ell;\theta\right),\label{e:eta_Z}\\
\tilde{\psi}_{Z}\left(\xb,\ell;\theta\right) & =\begin{cases}
\frac{\tilde{p}_{\sD}\left(\xb,\ell\right)\;g_k\left(\zb_{\theta\left(\ell\right)}|
\xb,\ell\right)}{\lambda c\left(\zb_{\theta\left(\ell\right)}\right)}, & \theta\left(\ell\right)>0\\
1-\tilde{p}_{{\sD}}\left(\mathbf{x},\ell\right), & \theta\left(\ell\right)=0
\end{cases}.\label{e:psi_Z}
\end{align}

In principle, the update involves generating a new $\delta$-GLMB
component for all possible associations of measurements to targets.
This is not usually feasible in practice, since the number of associations
increases combinatorially with the number of targets and measurements.
An effective method of reducing the number of posterior
$\delta$-GLMB components is to use a ranked assignment algorithm
to generate a requested number of components with highest weights.
For a set of targets with labels $\left\{ \ell_{1},\dots,\ell_{n}\right\} $,
and a set of measurements $\Zb=\{\zb_1,\dots,\zb_m\}$, this can
be achieved by constructing the following cost matrix,
\begin{equation}
-\log\left(\begin{array}{cccccc}
\vartheta^{\xi}\left(\ell_{1},\zb_{1}\right) & \cdots & \vartheta^{\xi}\left(\ell_{1},\zb_{m}\right) & \vartheta^{\xi}\left(\ell_{1},\emptyset\right) & \cdots & 0\\
\vdots & \ddots & \vdots & \vdots & \ddots & \vdots\\
\vartheta^{\xi}\left(\ell_{n},\zb_{1}\right) & \cdots & \vartheta^{\xi}\left(\ell_{n},\zb_{m}\right) & 0 & \cdots & \vartheta^{\xi}\left(\ell_{n},\emptyset\right)
\end{array}\right),\label{e:Cost_Matrix}
\end{equation}
where
\begin{align}
\vartheta^{\xi}\left(\ell,\zb\right) & =\sum_{i=1}^{N^{\xi}_{\ell}}\frac{w_{+}^{(i),\xi}\left(\ell\right)\tilde{p}_{{\sD}}
\left(\mathbf{x}_{+}^{(i),\xi},\ell\right)\,g_k\left(\zb|\mathbf{x}_{+}^{(i),\xi},
\ell\right)}{\lambda c\left(\zb\right)}\\
\vartheta^{\xi}\left(\ell,\emptyset\right) & =\sum_{i=1}^{N^{\xi}_{\ell}}w_{+}^{(i),\xi}\left(\ell\right)
\left(1-\tilde{p}_{{\sD}}\left(\mathbf{x}_{+}^{(i),\xi},\ell\right)\right)
\end{align}

In (\ref{e:Cost_Matrix}), each row represents a target, the first
$m$ columns represent detections, and the last $n$ columns represent
misdetections. Note that if measurement gating is carried out beforehand,
many of the detection terms in the first $m$ columns will be zero,
indicating that the corresponding measurement-to-target association
is infeasible. The matrix (\ref{e:Cost_Matrix}) is processed using
a ranked assignment algorithm. For example, Murty's algorithm \cite{murty_68},
yields a sorted list of the cheapest one-to-one assignments
of rows to columns, in increasing order of cost. Note that using this
formulation, each solution must assign every row to a column, but
there may be columns that are not assigned to a row. In generating
the list of ranked assignments, whenever a row is assigned to a column
index greater than $m$ by Murty's algorithm, the assignment is reported
as $0$, to indicate that the target was misdetected. Each element
in the list is then used to construct a component in the posterior
$\delta$-GLMB density.

Pseudo-code for the $\delta$-GLMB particle filter update step is given in Alg.
\ref{a:GLMB_Update}. Similarly to the prediction, the input includes
the components of the prior $\delta$-GLMB, and the number of posterior
components $M^{h}$ for $h=1,\dots,H_+$ to generate for each
prior component.

\begin{algorithm}[htb]
\caption{The update step of the $\delta$-GLMB particle filter }
{ \footnotesize
\begin{algorithmic}[1]
\Function{$\delta$-GLMB Particle Filter Update}{}
\State Input: $\Zb_k$, $\left\{ L_+^h, \phi_+^h,\left\{\left\{
         w_{+}^{(i),h}(\ell), \xb_{+}^{(i),h}(\ell) \right\}_{1\leq i\leq N^h_+(\ell)}\right\}_{\ell\in L^h_+}, M^h \right\}_{1\leq h \leq H_+}$
\State $m = 1$
\For{$h=1,\dots,H_+$}
\State Compute the cost matrix $C_Z^{h}$ according to (\ref{e:Cost_Matrix})
\State $\left\{\left(\theta^{h,j},c^{h,j}\right)\right\}_{1\leq j \leq M^{h}}
        = \mathtt{RankedAssignments}\left(L_+^{h},C_Z^{h},M^{h}\right)$
  \For{$j=1,\dots,M^{h}$}
    \State $L_k^m = L_+^{h}$
    \State $\phi_k^m = \phi_+^{h} \exp\left(-c^{h,j}\right)$
    \For{$\ell \in L_k^m$}
      \State $\tilde{\xb}_{k}^{(i),m}(\ell) = \xb_+^{(i),h}(\ell), \quad i=1,\dots,N^h_+(\ell)$
      \State $\tilde{w}_{k}^{(i),m}(\ell) = w_+^{(i),h}(\ell) \, \tilde{\psi}_Z(\xb_+^{(i),h}(\ell),\ell;\theta^{h,j})
              , \quad i=1,\dots,N^h_+(\ell)$
      \State $w_{k}^{(i),m}(\ell) =
              \tilde{w}_{k}^{(i),m}(\ell) \diagup \sum_{j=1}^{N^h_+(\ell)} \tilde{w}_{k}^{(j),m}(\ell)$ \Comment (Normalisation)
\For {$i=1,\dots,N^h_{\ell}$} \Comment (Resampling) \State Select index
$j^i\in\{1,\dots,N^h_+(\ell)\}$ with probability $w_{k}^{(i),m}(\ell)$ \State
 $\xb^{(i),m}_k(\ell) = \tilde{\xb}_{k}^{(j^i),m}(\ell)$ \EndFor
    \EndFor
    \State $m = m + 1$
  \EndFor
\EndFor
\State $H_k = m - 1$
\State Normalise weights $\left\{\phi_k^m\right\}_{1\leq m \leq H_k}$
\State Output:
       $\left\{ L_k^m,\phi_k^m,
        \left\{
        \left\{w_{k}^{(i),m}(\ell),\xb_{k}^{(i),m}(\ell)\right\}_{1\leq i\leq N^m_{\ell}}
        \right\}_{\ell\in L_k^m}
        \right\}_{1\leq m \leq H_k}$
\EndFunction
\end{algorithmic}}
\label{a:GLMB_Update}
\end{algorithm}

\subsubsection*{Estimation of labelled target states}

The final step in the filtering recursion is to estimate the labelled
target states based on the posterior $\delta$-GLMB density. One method is to
approximate the posterior $\delta$-GLMB in the form of a labelled
multi-Bernoulli PDF, where each target label has an associated existence
probability. Assuming the posterior $\delta$-GLMB density is specified by line 24 in Alg.
\ref{a:GLMB_Update}, for each $\ell\in\cup_{m=1}^{H_k} L_k^m$ the existence probability is computed as
\begin{equation}
r_{k,\ell}  =  \sum_{m=1}^{H_k} \phi_k^m\; 1_{L_k^m}(\ell).
\end{equation}
The corresponding spatial PDF $p_{k,\ell}(\xb)$ is approximated by a weighted particle set obtained as a union:
\[ \cup_{m=1}^{H_k} \begin{cases} \emptyset, & \mbox{ if } \ell\not\in L_k^m\\
\left\{\phi_k^m\cdot w_k^{(i),m}(\ell),\xb_k^{(i),m}(\ell) \right\}_{1\leq i \leq N^m_{\ell}}, & \mbox{ if } \ell\in L_k^m
\end{cases}
\]
A threshold can then be applied to $r_{k,\ell}$ determine which
tracks are to be reported.

A simpler alternative method, which we use here,
is to first compute the posterior cardinality distribution of the $\delta$-GLMB as follows:
\[
\rho(n) = \sum\limits_{m=1}^{H_k} \phi_k^m \cdot \delta [n,\left|L_k^m\right|],\quad n=0,1,2,\dots
\]
The  maximum a posteriori estimate of the target cardinality is then \[\hat{n} = \arg\max\limits_{n} \left[\rho(n)\right].\] The next step is to select a component $m^*$ as the highest weighted $\delta$-GLMB component representing cardinality $\hat{n}$. The set of reported
target states consist of the labels and weighted
particle sets in the component $m^*$.

\subsection{Demonstration: Bearings-only multi-object tracking}

\begin{figure}[tbh]
\centerline{\includegraphics[height=6cm]{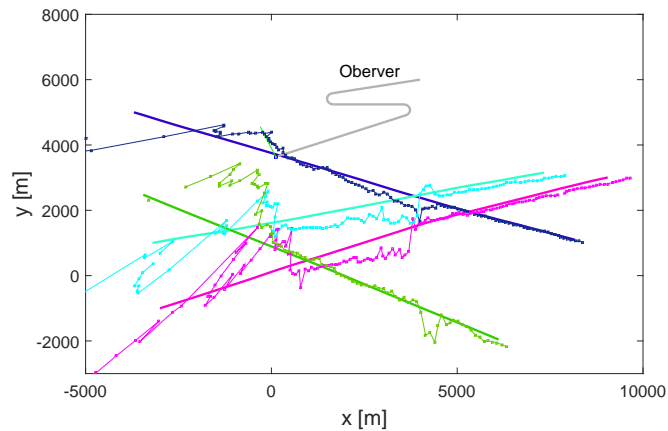}}
 \caption{A single run of the $\delta$-GLMB particle filter applied to bearings-only multi-target tracking.}
  \label{f:tracks}
\end{figure}

The performance of the described $\delta$-GLMB particle filter is
demonstrated by application to bearings-only multi-target tracking. The scenario
geometry and parameters are the same as those used in Sec. \ref{s:demo_phd}. The $\delta$-GLMB filter retains the top 100 hypotheses after each scan, and uses 5000 particles per target. The number of hypotheses generated in the prediction is 1000, and the number generated in the update is 4000. A single run output tracks of the $\delta$-GLMB particle filter are shown in Fig. \ref{f:tracks}. The ground truth tracks are shown by thick solid lines (compare with Fig. \ref{f:phd}.a), while the estimated tracks are represented by the same coloured thin lines. Because all four targets in the scenario are moving from left to right, the accuracy of estimated tracks also improves from left to right.

\begin{figure}[tbh]
\centerline{\includegraphics[height=6cm]{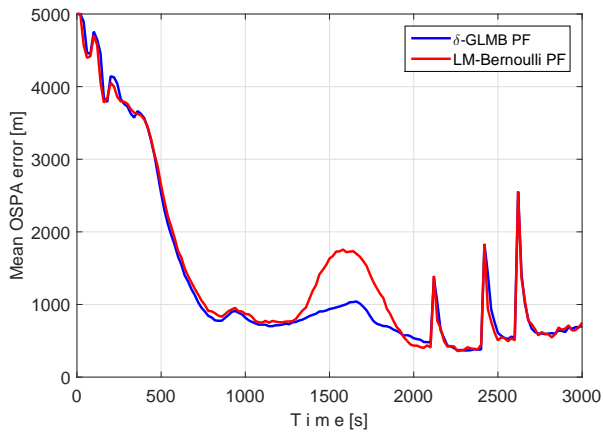}}
 \caption{Bearings-only multi-target tracking results: the mean OSPA error of the $\delta$-GLMB particle filter (blue) and the LM-Bernoulli particle filter (red).}
  \label{f:bo_mtt}
\end{figure}

The average tracking performance is studied next.  For comparison sake, we also present the tracking results of the Bernoulli particle filter (Sec. \ref{s:bpf})  which has been modified
to track multiple targets via the use of the ``linear multi-target''
(LM) technique proposed in \cite{musicki_lm_08}. The LM technique is an approximation
in which the detections from the nearby targets are considered as false detections that affect only
the clutter density $c(\zb)$.  This filter is referred
to as the LM-Bernoulli particle filter. Fig. \ref{f:bo_mtt} shows the mean OSPA errors, obtained by averaging over 500 Monte Carlo runs of the $\delta$-GLMB PF and the LM-Bernoulli PF. One can observe that
the average performance of the two tracking filters is almost equivalent when
the targets are well separated. However, when the targets approach each other (from around 1400s to 1900s), the $\delta$-GLMB PF exhibits better performance. This is to be expected since the $\delta$-GLMB PF  is an implementation of the principled Bayes-optimal multi-target tracker.
We can also comment on the performance of the two Bernoulli-type tracking particle filters versus the PHD particle filters, whose OSPA errors were shown in Fig. \ref{f:phd}.(b). First, recall that the tracking filters produce tracks (as opposed to the PHD filters, which produce only multi-target state estimates).  Comparing Figs. \ref{f:phd} and \ref{f:bo_mtt} we also observe that both Bernoulli-type tracking particle filters achieve somewhat lower mean OSPA errors than the PHD particle filter described in Sec. \ref{s:phd}.

%

%

\section{Summary and further reading}
\label{s:sum}

This overview paper is an attempt to describe the particle method based implementation of the new class of Bayes-optimal and suboptimal filters, derived using random finite set models. The primary focus was on the Bernoulli PF, the probability hypothesis density PF and the $\delta$-generalised labelled multi-Bernoulli PF. Each of these particle filters has been studied in detail and subsequently demonstrated in the context of bearings-only filtering/tracking.

Particle filters for random finite set models is a growing research field and there are many topics that this overview paper could not address due to the space limitation. For an interested reader we briefly list some of these topics: different (non-standard) measurement models (e.g. intensity measurements \cite{vo_tbd}, \cite{papi_15}, extended versus point target measurements \cite{granstrom2012extended},\cite{ristic2013bernoulli},\cite{liu2015sequential},\cite{beard_16}, group tracking
\cite{swain_clark_11}, finite sensor resolution or merged measurements \cite{mahler_unresolved}, \cite{lian_12},\cite{beard_15}); multi-sensor fusion aspects \cite{uney2010monte}; smoothing (as opposed to filtering) \cite{mahler_phd_sm}; sensor control aspects \cite{automatica_10} and handling the switching dynamic models \cite{dunne2013multiple}.



\small


\begin{thebibliography}{10}
\expandafter\ifx\csname url\endcsname\relax
  \def\url#1{\texttt{#1}}\fi
\expandafter\ifx\csname urlprefix\endcsname\relax\def\urlprefix{URL
}\fi \expandafter\ifx\csname href\endcsname\relax
  \def\href#1#2{#2} \def\path#1{#1}\fi

\bibitem{jazwinski_70}
A.~H. Jazwinski.
\newblock {\em Stochastic processes and filtering theory}.
\newblock Academic press, 1970.

\bibitem{smcbook}
A.~Doucet, J.~F.~G. de~Freitas, and N.~J. Gordon, editors.
\newblock {\em Sequential Monte Carlo Methods in Practice}.
\newblock Springer, 2001.

\bibitem{pf_tute}
M.~S. Arulampalam, S.~Maskell, N.~Gordon, and T.~Clapp.
\newblock A tutorial on particle filters for non-linear/non-{G}aussian
  {B}ayesian tracking.
\newblock {\em IEEE Trans. Signal Processing}, 50(2):174--188, Feb. 2002.

\bibitem{djuric_pf_03}
P.~Djuric, J.~H. Kotecha, J.~Zhang, Y.~Huang, T.~Ghirmai, M.~Bugallo, and
  J.~Miguez.
\newblock Particle filtering.
\newblock {\em IEEE Signal Processing Magazine}, pages 19--38, Sept. 2003.

\bibitem{pfbook}
B.~Ristic, S.~Arulampalam, and N.~Gordon.
\newblock {\em Beyond the {K}alman filter: {P}article filters for tracking
  applications}.
\newblock Artech House, 2004.

\bibitem{cape_07}
O.~Capp\'{e}, S.~J. Godsill, and E.~Moulines.
\newblock An overview of existing methods and recent advances in sequential
  {M}onte {C}arlo.
\newblock {\em Proc. IEEE}, 95(5):899--924, 2007.

\bibitem{doucet_15_years}
A.~Doucet and A.~M. Johansen.
\newblock A tutorial on particle filtering and smoothing: Fifteen years later.
\newblock {\em Handbook of Nonlinear Filtering}, 12:656--704, 2009.

\bibitem{kay1998}
S.~M. Kay.
\newblock Fundamentals of statistical signal processing, {V}ol. {II}:
  {D}etection theory.
\newblock {\em Prentice Hall}, 1998.

\bibitem{mahler_14}
R.~P.~S. Mahler.
\newblock {\em Advances in Statisical Multisource-multitarget information
  fusion}.
\newblock Artech House, 2014.

\bibitem{barshalom_et_al_01}
Y.~Bar-Shalom, X.~R. Li, and T.~Kirubarajan.
\newblock {\em Estimation with Applications to Tracking and Navigation}.
\newblock John Wiley \& Sons, 2001.

\bibitem{julier2004unscented}
S.~J. Julier and J.~K. Uhlmann.
\newblock Unscented filtering and nonlinear estimation.
\newblock {\em Proceedings of the IEEE}, 92(3):401--422, 2004.

\bibitem{challa}
S.~Challa, M.~R. Morelande, D.~Mu\v{s}icki, and R.~J. Evans.
\newblock {\em Fundamentals of Object Tracking}.
\newblock Cambridge Univ. Press, 2011.

\bibitem{robert_casella99}
C.~P. Robert and G.~Casella.
\newblock {\em Monte Carlo statistical methods}.
\newblock Springer, 2nd edition, 2004.

\bibitem{gilks}
C.~Berzuini and W.~R. Gilks.
\newblock Resample-move filtering with cross-model jumps.
\newblock In A.~Doucet, N.~De Freitas, and N.~Gordon, editors, {\em Sequential
  Monte Carlo Methods in Practice}, chapter~6. Springer, 2001.

\bibitem{gordon_etal_93}
N.~J. Gordon, D.~J. Salmond, and A.~F.~M. Smith.
\newblock Novel approach to nonlinear/non-{G}aussian {B}ayesian state
  estimation.
\newblock {\em IEE Proc.-F}, 140(2):107--113, 1993.

\bibitem{doucetetal99}
A.~Doucet, S.~Godsill, and C.~Andrieu.
\newblock On sequential {M}onte {C}arlo sampling methods for {B}ayesian
  filtering.
\newblock {\em Statistics and Computing}, 10(3):197--208, 2000.

\bibitem{PITTSHEPHARD}
M.~Pitt and N.~Shephard.
\newblock Filtering via simulation: Auxiliary particle filters.
\newblock {\em Journal of the American Statistical Association},
  94(446):590--599, 1999.

\bibitem{merwe}
R.~Van der Merwe, A.~Doucet, N.~De Freitas, and E.~Wan.
\newblock The unscented particle filter.
\newblock In {\em Advances in Neural Information Processing Systems},
  volume~13. 2000.

\bibitem{daum_10}
F.~Daum, J.~Huang, and A.~Noushin.
\newblock Exact particle flow for nonlinear filters.
\newblock In {\em Proc. SPIE}, volume 7697, 2010.

\bibitem{musso_et_al_00}
C.~Musso, N.~Oudjane, and F.~LeGland.
\newblock Improving regularised particle filters.
\newblock In A.~Doucet, N.~deFreitas, and N.~J. Gordon, editors, {\em
  Sequential {M}onte {C}arlo methods in {P}ractice}. Springer, 2001.

\bibitem{paul_quang_15}
P.~B. Quang, C.~Musso, and F.~Le Gland.
\newblock Particle filtering and the laplace method for target tracking.
\newblock {\em IEEE Trans. Aerospace and Electronic Systems}, 2015.
\newblock (to appear).

\bibitem{silverman_86}
B.~W. Silverman.
\newblock {\em Density estimation for statistical and data analysis}.
\newblock Chapman and Hall, 1986.

\bibitem{pMCMC}
C.~Andrieu, A.~Doucet, and R.~Holenstein.
\newblock Particle {M}arkov chain {M}onte {C}arlo methods.
\newblock {\em Journ. Royal Statistical Soc. B}, 72(Part 3):269--342, 2010.

\bibitem{schon_05}
T.~Schon, F.~Gustafsson, and P.~J. Nordlund.
\newblock Marginalized particle filters for mixed linear/nonlinear state-space
  models.
\newblock {\em IEEE Trans Signal Processing}, 53(7):2279--2289, 2005.

\bibitem{SRF_07}
S.~Arulampalam, M.~Clark, and R.~Vinter.
\newblock Performance of the shifted {R}ayleigh filter in single-sensor
  bearings-only tracking.
\newblock In {\em Proc. 10th IEEE Int. Conf. Information Fusion}, July 2007.

\bibitem{beard_15}
M.~Beard, B.-T. Vo, and B.-N. Vo.
\newblock Bayesian multi-target tracking with merged measurements using
  labelled random finite sets.
\newblock {\em IEEE Trans. Signal Processing}, 63(6):1433--1447, 2015.

\bibitem{mahler_07}
R.~Mahler.
\newblock {\em Statistical Multisource Multitarget Information Fusion}.
\newblock Artech House, 2007.

\bibitem{vo_05}
B.~N. Vo, S.~Singh, and A.~Doucet.
\newblock Sequential {M}onte {C}arlo methods for multi-target filtering with
  random finite sets.
\newblock {\em IEEE Trans. Aerospace \& Electronic Systems}, 41(4):1224--1245,
  Oct. 2005.

\bibitem{bf_tute}
B.~Ristic, B.-T. Vo, B.-N. Vo, and A.~Farina.
\newblock A tutorial on {B}ernoulli filters: {T}heory, implementation and
  applications.
\newblock {\em IEEE Trans. Signal Processing}, 61(13):3406--3430, 2013.

\bibitem{springer_book_13}
B.~Ristic.
\newblock {\em Particle filters for random set models}.
\newblock Springer, 2013.

\bibitem{boers_04}
Y.~Boers and J.~N. Driessen.
\newblock A particle filter multi target track before detect application.
\newblock {\em IEE Proc. Radar, Sonar and Navigation}, 151(6):351--357, 2004.

\bibitem{sidenbladh_03}
H.~Sidenbladh and S.~L. Wirkander.
\newblock Tracking random sets of vehicles in terrain.
\newblock In {\em Proc. 2nd IEEE Workshop on Multi-Object Tracking}, Madison,
  WI, USA, June 2003.

\bibitem{zajic_03}
T.~Zajic and R.~Mahler.
\newblock A particle-systems implementation of the {PHD} multitarget tracking
  filter.
\newblock In {\em Proc. SPIE}, volume 5096, pages 291--299, April 2003.

\bibitem{vihola_07}
M.~Vihola.
\newblock Rao-{B}lackwellised particle filtering in random set multitarget
  tracking.
\newblock {\em IEEE Trans. Aerospace \& Electronic Systems}, 43(2):689--705,
  2007.

\bibitem{automatica_10}
B.~Ristic and B.-N. Vo.
\newblock Sensor control for multi-object state-space estimation using random
  finite sets.
\newblock {\em Automatica}, 46:1812--1818, 2010.

\bibitem{reuter_11}
S.~Reuter and K.~Dietmayer.
\newblock Pedestrian tracking using random finite sets.
\newblock In {\em Proc. Int. Conf. Information Fusion}, Chicago, USA, July
  2011.

\bibitem{boers_mixed_labeling}
Y.~Boers, E.~Sviestins, and H.~Driessen.
\newblock Mixed labelling in multitarget particle filtering.
\newblock {\em IEEE Trans Aerospace and Electronic Systems}, 46(2):792--802,
  2010.

\bibitem{mahler_phdf_03}
R.~P.~S. Mahler.
\newblock Multi-target {B}ayes filtering via first-order multi-target moments.
\newblock {\em IEEE Trans. Aerospace \& Electronic Systems}, 39(4):1152--1178,
  2003.

\bibitem{tobias_05}
M.~Tobias and A.D. Lanterman.
\newblock Probability hypothesis density-based multitarget tracking with
  bistatic range and {D}oppler observations.
\newblock {\em IEE Proc.-Radar Sonar Navig}, 152(3):195--205, 2005.

\bibitem{clark_sonar}
D.~Clark, I.~T. Ruiz, Y.~Petillot, and J.~Bell.
\newblock Particle {PHD} filter multiple target tracking in sonar image.
\newblock {\em IEEE Trans. Aerospace \& Electronic Systems}, 43(1):409--416,
  2007.

\bibitem{maggio_08}
E.~Maggio, M.~Taj, and A.~Cavallaro.
\newblock Efficient multitarget visual tracking using random finite sets.
\newblock {\em IEEE Trans. Circuits \& Systems for Video Technology},
  18(8):1016--1027, 2008.

\bibitem{lundquist2011road}
C.~Lundquist, L.~Hammarstrand, and F.~Gustafsson.
\newblock Road intensity based mapping using radar measurements with a
  probability hypothesis density filter.
\newblock {\em IEEE Trans. on Signal Processing}, 59(4):1397--1408, 2011.

\bibitem{canaud2013probabilty}
M.~Canaud, L.~Mihaylova, J.~Sau, and N.-E.~El Faouzi.
\newblock Probabilty hypothesis density filtering for real-time traffic state
  estimation and prediction.
\newblock {\em Networks and Heterogeneous Media (NHM)}, 8(3):825--842, 2013.


\bibitem{mullane_11}
 J. ~Mullane, B.-N. Vo, M. D. Adams and B.-T. Vo.
\newblock A Random-Finite-Set Approach to Bayesian SLAM.
\newblock {\em IEEE Transactions on Robotics},
27(2):268-282, 2011.


\bibitem{adams2014slam}
M.~Adams, B.-N. Vo, R.~Mahler, and J.~Mullane.
\newblock {SLAM} gets a {PHD}: {N}ew concepts in map estimation.
\newblock {\em IEEE Robotics \& Automation Magazine}, 21(2):26--37, 2014.

\bibitem{lee2014slam}
C.~S. Lee, S.~Nagappa, N.~Palomeras, D.~E. Clark, and J.~Salvi.
\newblock Slam with {SC-PHD} filters: {A}n underwater vehicle application.
\newblock {\em IEEE Robotics \& Automation Magazine}, 21(2):38--45, 2014.

\bibitem{rezatofighi_13}
S.~H. Rezatofighi, S.~Gould, B.-N. Vo, K.~Mele, and W.~E. Hughesand~R. Hartley.
\newblock A multiple model probability hypothesis density tracker for
  time-lapse cell microscopy sequences.
\newblock In {\em Information Processing in Medical Imaging}, pages 110--122.
  Springer, 2013.

\bibitem{mahler_cphdf_07}
R.~P.~S. Mahler.
\newblock {PHD} filters of higher order in target number.
\newblock {\em IEEE Trans. Aerospace \& Electronic Systems}, 43(4):1523--1543,
  2007.

\bibitem{ristic2012adaptive}
B.~Ristic, D.~Clark, B.-N. Vo, and B.-T. Vo.
\newblock Adaptive target birth intensity for {PHD} and {CPHD} filters.
\newblock {\em IEEE Trans. on Aerospace and Electronic Systems},
  48(2):1656--1668, 2012.

\bibitem{Whiteley_aes}
N.~P. Whiteley, S.~S. Singh, and S.~J. Godsill.
\newblock Auxiliary particle implementation of the probability hypothesis
  density filter.
\newblock {\em IEEE Trans. on Aerospace \& Electronic Systems},
  46(3):1437--1454, July 2010.

\bibitem{yoon_12}
J.~H. Yoon, D.~Y. Kim, and K.-Y. Yoon.
\newblock Efficient importance sampling function design for sequential {M}onte
  {C}arlo {PHD} filter.
\newblock {\em Signal Processing}, 92:2315--2321, 2012.

\bibitem{high_speed_13}
T.~Li, S.~Sun, and T.~P. Sattar.
\newblock High speed sigma-gating {SMC}-{PHD} filter.
\newblock {\em Signal Processing}, 93:2586--2593, 2013.

\bibitem{ristic_icassp_15}
B.~Ristic.
\newblock Efficient update of persistent particles in the {SMC-PHD} filter.
\newblock In {\em Proc. IEEE Int. Conf. Acoustic Speech Signal Processing
  (ICASSP)}, Brisbane, Australia, 2015.

\bibitem{clark_05}
D.~E. Clark, J.~Bell, Y.~de~Saint-Pern, and Y.~Petillot.
\newblock {PHD} filter for multi-target tracking in {3D} sonar.
\newblock In {\em Proc. IEEE OCEANS-05-Europe}, pages 265--270, Brest, France,
  2005.

\bibitem{calib_13}
B.~Ristic, D.~E. Clark, and N.~Gordon.
\newblock Calibration of multi-atrget tracking algorithms using non-cooperative
  targets.
\newblock {\em IEEE Journal of Selected Topics in Signal Processing},
  7(3):390--398, 2013.

\bibitem{icassp_12_sr}
B.~Ristic and D.~Clark.
\newblock Particle filter for joint estimation of multi-object dynamic state
  and multi-sensor bias.
\newblock In {\em Proc. IEEE Int. Conf. Acoustics, Speech \& Signal Proc.
  (ICASSP)}, pages 3877--3880, Kyoto, Japan, March 2012.

\bibitem{swainspie10}
A.~Swain and D.~E. Clark.
\newblock {First-moment filters for spatial independent cluster processes}.
\newblock {\em Proc. of SPIE 7697}, 2010.

\bibitem{swainfusion10}
A.~Swain and D.~Clark.
\newblock {Extended object filtering using spatial independent cluster
  processes}.
\newblock {\em Proc. Int. Conf. Information Fusion}, 2010.

\bibitem{lee2013slam}
C.~S. Lee, D.~E. Clark, and J.~Salvi.
\newblock {SLAM} with dynamic targets via single-cluster {PHD} filtering.
\newblock {\em IEEE Journal of Selected Topics in Signal Processing,},
  7(3):543--552, 2013.

\bibitem{SchuhmacherOSPA07}
D.~Schuhmacher, B.-T. Vo, and B.-N. Vo.
\newblock A consistent metric for performance evaluation of multi-object
  filters.
\newblock {\em IEEE Trans. Signal Processing}, 56(8):3447--3457, 2008.

\bibitem{lin_06}
L.~Lin, Y.~bar Shalom, and T.~Kirubarajan.
\newblock Track labelling and {PHD} filter for multitarget tracking.
\newblock {\em IEEE Trans Aerospace and Electronic Systems}, 42(3):778--795,
  2006.

\bibitem{yang_12}
J.~Yang and H.~Ji.
\newblock A novel track maintenance algorithm for {PHD/CPHD} filter.
\newblock {\em Signal Processing}, 92:2371--2380, 2012.

\bibitem{cbmember_09}
B.-T. Vo, B.~N. Vo, and A.~Cantoni.
\newblock The cardinality balanced multi-target multi-{B}ernoulli filter and
  its implementations.
\newblock {\em IEEE Trans. on Signal Processing}, 57(2):409--423, 2009.

\bibitem{vo2013labeled}
B.-T. Vo and B.-N. Vo.
\newblock Labeled random finite sets and multi-object conjugate priors.
\newblock {\em IEEE Trans. on Signal Processing}, 61(13):3460--3475, 2013.

\bibitem{vo2014labeled}
B.-N. Vo, B.-T. Vo, and D.~Phung.
\newblock Labeled random finite sets and the {B}ayes multi-target tracking
  filter.
\newblock {\em IEEE Trans. on Signal Processing}, 62(24):6554--6567, 2014.

\bibitem{williams_15}
J. L. Williams.
\newblock An efficient, variational approximation of the best fitting multi-{B}ernoulli filter.
\newblock {\em IEEE Trans. Signal Processing}, 63(1):258--273, 2015.

\bibitem{correa_15}
J. Correa, M. Adams, and C. Perez.
 \newblock A Dirac delta mixture-based random finite set filter.
 \newblock {\em Proc. IEEE Int. Conf. Control, Automation and Information Sciences (ICCAIS)}, 2015.


\bibitem{murty_68}
K.~G. Murty.
\newblock An algorithm for ranking all the assignments in order of increasing
  cost.
\newblock {\em Operations Research}, 16(3):682—687, 1968.

\bibitem{musicki_lm_08}
D.~Musicki and B.~La Scala.
\newblock Multi-target tracking in clutter without measurement assignment.
\newblock {\em IEEE Trans. Aerospace and Electronic Systems}, 44(3):877--896,
  July 2008.

\bibitem{vo_tbd}
B.-N. Vo, B.-T. Vo, N.-T. Pham, and D.~Suter.
\newblock Joint detection and estimation of multiple objects from image
  observations.
\newblock {\em IEEE Trans. Signal Processing}, 58(10):5129--5141, 2010.

\bibitem{papi_15}
F.~Papi, B.-N. Vo, B.-T. Vo, C.~Fantacci, and M.~Beard.
\newblock Generalized labeled multi-{B}ernoulli approximation of multi-object
  densities.
\newblock {\em IEEE Trans. Signal Processing}, 63(20):5487--5497, Oct. 2015.

\bibitem{granstrom2012extended}
K.~Granstr{\"o}m, C.~Lundquist, and O.~Orguner.
\newblock Extended target tracking using a {G}aussian-mixture {PHD} filter.
\newblock {\em IEEE Trans. on Aerospace and Electronic Systems},
  48(4):3268--3286, 2012.

\bibitem{ristic2013bernoulli}
B.~Ristic and J.~Sherrah.
\newblock Bernoulli filter for joint detection and tracking of an extended
  object in clutter.
\newblock {\em IET Radar, Sonar \& Navigation}, 7(1):26--35, 2013.

\bibitem{liu2015sequential}
Meiqin Liu, Tongyang Jiang, and Senlin Zhang.
\newblock The sequential {M}onte {C}arlo multi-{B}ernoulli filter for extended
  targets.
\newblock In {\em 18th Intern. Conf. on Information Fusion}, pages 984--990,
  2015.

\bibitem{beard_16}
M.~Beard, S.~Reuter, K.~Granstr\"{o}m, B.-T. Vo, B.-N. Vo, and A.~Scheel.
\newblock Multiple extended target tracking with labelled random finite sets.
\newblock {\em IEEE Trans. Signal Processing}, 2016.
\newblock In print.

\bibitem{swain_clark_11}
A.~Swain and D.~Clark.
\newblock The single-group {PHD} filter: an analytic solution.
\newblock In {\em Proc. 14th Int. Conf. Information Fusion}, Chicago, USA, July
  2011.

\bibitem{mahler_unresolved}
R.~Mahler.
\newblock {PHD} filters for nonstandard targets, {II}: unresolved targets.
\newblock In {\em Proc. 12th Int. Conf. Information Fusion}, pages 922–--929,
  Seattle, USA, 2009.

\bibitem{lian_12}
F.~Lian, C.~Han, W.~Liu, J.~Liu, and J.~Sun.
\newblock Unified cardinalized probability hypothesis density filters for
  extended targets and unresolved targets.
\newblock {\em Signal Processing}, 92(7):1729--1744, 2012.

\bibitem{uney2010monte}
M.~Uney, S.~Julier, D.~Clark, and B.~Ristic.
\newblock Monte carlo realisation of a distributed multi-object fusion
  algorithm.
\newblock In {\em IET Sensor Signal Processing for Defence (SSPD 2010)}, 2010.

\bibitem{mahler_phd_sm}
R.~P.~S. Mahler, B.-T. Vo, and B.-N. Vo.
\newblock Forward-backward probability hypothesis density smoothing.
\newblock {\em IEEE Trans. Aerospace and Electronic Systems}, 48(1):707--728,
  Jan 2012.

\bibitem{dunne2013multiple}
D.~Dunne and T.~Kirubarajan.
\newblock Multiple model multi-{B}ernoulli filters for manoeuvering targets.
\newblock {\em IEEE Trans. Aerospace and Electronic Systems}, 49(4):2679--2692,
  2013.

\end{thebibliography}

\end{document}